\documentclass[preprint,12pt]{elsarticle}

\usepackage{graphics}
\usepackage{color}
\definecolor{rouge}{rgb}{1,0,0}
\definecolor{bleu}{rgb}{0,0,1}
\definecolor{vert}{rgb}{0,0.5,0}

\usepackage{amssymb}
\usepackage{cases}

\journal{Wave Motion}

\begin{document}

\begin{frontmatter}

\title{Semi-analytical and numerical methods for computing transient waves in 2D acoustic / poroelastic stratified media}

\author[Avignon]{G. Lefeuve-Mesgouez\corref{cor1}}
\ead{gaelle.mesgouez@univ-avignon.fr}
\author[Avignon]{A. Mesgouez\corref{cor2}}
\ead{arnaud.mesgouez@univ-avignon.fr}
\author[M2P2]{G. Chiavassa\corref{cor3}}
\ead{guillaume.chiavassa@centrale-marseille.fr}
\author[LMA]{B. Lombard\corref{cor4}}
\ead{lombard@lma.cnrs-mrs.fr}

\cortext[cor1]{Corresponding author}

\address[Avignon]{Universit\'e d'Avignon et des Pays de Vaucluse, UMR EMMAH, Facult\'e des Sciences, 33 rue Louis Pasteur, F-84000 Avignon, France}
\address[M2P2]{Laboratoire de M\'ecanique, Mod\'elisation et Proc\'ed\'es Propres, UMR 6181 CNRS-ECM, Technop\^ole de Chateau-Gombert, 38 rue Joliot-Curie, 13451 Marseille, France}
\address[LMA]{Laboratoire de M\'ecanique et d'Acoustique, UPR 7051 CNRS, 31 chemin Joseph Aiguier, 13402 Marseille, France}

\begin{abstract}
Wave propagation in a stratified fluid / porous medium is studied here using analytical and numerical methods. The semi-analytical method is based on an exact stiffness matrix method coupled with a matrix conditioning procedure, preventing the occurrence of poorly conditioned numerical systems. Special attention is paid to calculating the Fourier integrals. The numerical method is based on a high order finite-difference time-domain scheme. Mesh refinement is applied near the interfaces to discretize the slow compressional diffusive wave predicted by Biot's theory. Lastly, an immersed interface method is used to discretize the boundary conditions. The numerical benchmarks are based on realistic soil parameters and on various degrees of hydraulic contact at the fluid / porous boundary. The time evolution of the acoustic pressure and the porous velocity is plotted in the case of one and four interfaces. The excellent level of agreement found to exist between the two approaches confirms the validity of both methods, which cross-checks them and provides useful tools for future researches.
\end{abstract}


\end{frontmatter}


\section{Introduction}\label{SecIntro}

This study focuses on the propagation of transient 2D mechanical waves emitted by a source point in a fluid half-space over a stratified poroelastic medium. This configuration occurs in situations such as those encountered in the field of underwater acoustics and civil engineering. 
Natural or artificial media, presenting unidirectional varying properties, can be modelled as multilayered structures.
The wave propagation is  described by means of acoustic equations in the fluid and low-frequency Biot's equations in the poroelastic media \cite{BIOT56-A}. Hydraulic exchanges between the fluid and the first porous layer are modeled using various boundary conditions \cite{ROSENBAUM74,GUREVICH99}. Classical textbooks such as \cite{BOURBIE87,CARCIONE07} can be consulted for a detailed analysis of Biot's equations (which involve a fast compressional wave and a shear wave, as in elastic media, and a slow compressional wave). In the case of a single fluid / porous interface, many theoretical studies have dealt with the reflection / transmission coefficients \cite{WU90} and with the properties of the surface waves \cite{FENG83a,DENNEMAN02,EDELMAN04,GUBAIDULLIN04,DALEN10}. The aim of the present study is to solve this problem in the case of an arbitrary number of layers, using two radically different approaches: a semi-analytical approach and a purely numerical one.  

On the one hand, various analytical methods have been developed in the case of a single fluid / porous interface:  Cagniard-de Hoop's method \cite{FENG83b,DIAZ10} for an inviscid medium, and \cite{LU07,DALEN11} for a viscous saturating fluid, to cite but a few. In situations involving a larger number of interfaces, the strategies usually adopted are based on transfer matrix, stiffness matrix, or transmission and reflection matrix methods. These approaches were first developed for use with viscoelastic media \cite{THOMSON50,HASKELL53,KAUSEL81}, and only a few studies using multilayer approaches to poroelastic models have been published so far because of the complexity of poroelastic models and the poor conditioning of the remaining matrices. The exact stiffness matrix method has been applied in the context of poroelasticity to harmonic 2D problems \cite{RAJAPAKSE95}, transient 2D problems \cite{DEGRANDE98}, and problems involving axisymmetric quasi-static configurations \cite{SENJUNTICHAI95}. All the latter authors stressed the numerical difficulties encountered due to poorly conditioned matrices. Methods have been proposed in \cite{LU05b} for eliminating the mismatched terms using the transmission and reflexion matrix approach in the case of axisymmetric geometries, and recently, using the exact stiffness matrix method in \cite{MESGOUEZ09}. This method is extended here to the coupling between fluid and porous media by simulating the various interface conditions pertaining at the boundary between the fluid and the stratified porous media.

On the other hand, only a few numerical methods have addressed realistic acoustic / poroelastic configurations: for this purpose, a spectral-element method \cite{MORENCY08}, a pseudospectral method \cite{SIDLER10} and a discontinuous Galerkin method \cite{DUPUY_THESE} have been developed. Three difficulties have to be overcome to be able to perform reliable and efficient numerical simulations. First, the slow compressional Biot wave is highly attenuated, which drastically affects the stability  of any explicit numerical scheme \cite{PUENTE08}. Secondly, the slow compressional wave is always located near the interfaces, and thus plays a key role in the balance equations; discretizing this small-scale wave therefore requires a particularly fine spatial mesh. Thirdly, large numerical errors are usually introduced by the boundaries due to the poor discretization of their geometrical and physical properties, and also due to issues involved in the numerical analysis. In the present study, it is proposed to adopt a strategy we have described in previous articles \cite{CHIAVASSA11,CHIAVASSA12}. A fourth-order finite-difference scheme with splitting is used to integrate the evolution equations, so as to optimize the integration time step. Specific solvers are used in the case of the fluid and the porous media. Their coupling is obtained using an immersed interface method to discretize the boundary conditions, which gives a subcell resolution of the geometries. Lastly, a space-time mesh refinement procedure is applied around the interfaces to account for the small scale of the slow compressional wave. 

The paper is organized as follows. In section \ref{SecPS}, notations, governing equations and boundary conditions are stated. In section \ref{SecAnal}, the semi-analytical approach is described, focusing on the coupling between fluid and porous media and on the integration of oscillating functions. In section \ref{SecNum}, the key components of the numerical approach are recalled. In section \ref{SecRes}, numerical experiments are performed on situations involving one and four interfaces, with realistic soil parameters and under the low-frequency Biot theory. Comparisons between the analytical and numerical solutions obtained confirm the validity of both approaches. From the numerical point of view, this study provides a useful benchmark for future studies on viscous saturating fluids that have not been available so far \cite{MORENCY08,SIDLER10,DUPUY_THESE}. From the semi-analytical point of view, these comparisons also show the relevance of using the chosen integration parameters, for which no theoretical criterion exists. Concluding remarks and some future lines of research are presented in the last section \ref{SecConclu}.

\section{Statement of the problem}\label{SecPS}

\subsection{Governing equations}\label{SecPSedp}

\begin{figure}[htbp]
\begin{center}
\begin{tabular}{c}
\includegraphics[scale=0.70]{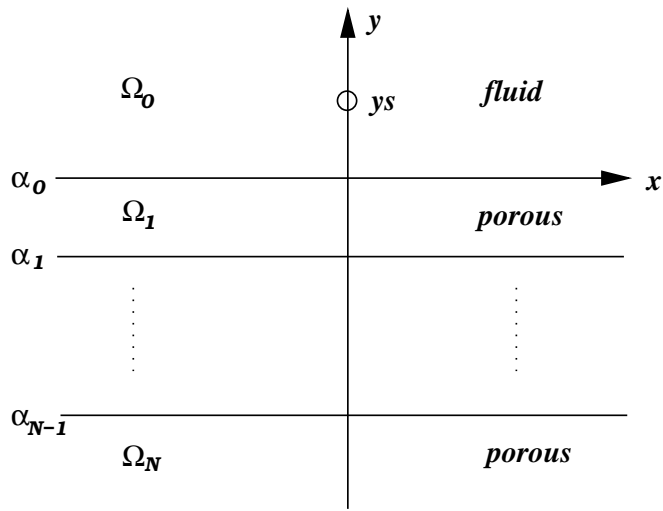} 
\end{tabular}
\end{center}
\caption{Configuration under study. Fluid medium over a multilayered porous medium. Source point $y_s$ in the fluid.}
\label{FigCrobar}
\end{figure}

The 2D configuration under investigation is a fluid half-space $\Omega_0$ over a stack of homogeneous poroelastic layers $\Omega_j$ ($j=1,\cdots, N$), as shown in Fig. \ref{FigCrobar}. The $x$ and $y$ geometrical axes point rightward and upward, respectively. The $N$ plane and parallel interfaces are located at $y=\alpha_i\leq 0$, where $\alpha_0=0$. A source point at $(x_s=0,\, y_s>0)$ in the fluid emits cylindrical waves. 

In the fluid domain $\Omega_0$, the physical parameters are the density $\rho_f$ and the celerity of acoustic waves $c$. The acoustic equations are written as follows
\begin{equation}
\left\{
\begin{array}{l}
\displaystyle
p=-\rho\,c^2\,{\bf \nabla.U},\\
[8pt]
\displaystyle
\Delta\,p-\frac{\textstyle 1}{\textstyle c^2}\,\ddot{p}=-S(t)\,\delta(x)\,\delta(y-y_s),
\end{array}
\right.
\label{LCAcous}
\end{equation}
where ${\bf U}=(U_x,\,U_y)^t$ is the fluid displacement, $p$ is the acoustic pressure, and $S(t)$ is a causal source term. The overlying dot denotes the derivative in terms of time $t$.

The poroelastic media $\Omega_j$ ($j=1,\cdots, N$) are modeled using the low-frequency Biot theory \cite{BIOT56-A,BOURBIE87}. The physical parameters are 
\begin{itemize}
\item the dynamic viscosity $\eta$ and the density $\rho_f$ of the saturating fluid. The latter is assumed to be the same as in $\Omega_0$, and the notation $\rho_f$ is therefore used in both cases; 
\item the density $\rho_s$ and the shear modulus $\mu$ of the elastic skeleton; 
\item the connected porosity $\phi$, the tortuosity $a_{\infty}$, the absolute permeability $\kappa$, the Lam\'e coefficient of the dry matrix $\lambda_0$, and the two Biot coefficients $\beta$ and $m$ of the isotropic matrix. 
\end{itemize}
We thus obtain the total density $\rho=\phi\,\rho_f+(1-\phi)\,\rho_s$. The low-frequency Biot model is valid at frequencies below the critical value defined as 
\begin{equation}
f_c=\frac{\textstyle \eta\,\phi}{\textstyle 2\,\pi\,a_\infty\,\kappa\,\rho_f}.
\label{Fc}
\end{equation}
Based on the constitutive equations and the conservation of momentum in porous media, one obtains \cite{BIOT56-A}
\begin{equation}
\left\{
\begin{array}{l}
\displaystyle
{\bf \sigma}=\left(\lambda_0\,{\bf \nabla.u}-\beta\,p\right)\,{\bf I}+2\,\mu\,{\bf \varepsilon},\\
[10pt]
\displaystyle
p=-m\left(\beta\,{\bf \nabla.u}+{\bf \nabla.w}\right),\\
[10pt]
\displaystyle
{\bf \nabla\,\sigma}=\rho\,\ddot{\bf u}+\rho_f\,\ddot{\bf w},\\
[10pt]
\displaystyle
-{\bf \nabla}\,p=\rho_f\,\ddot{\bf u}+\frac{\textstyle a_{\infty}\,\rho_f}{\textstyle \phi}\,\ddot{\bf w}+\frac{\textstyle \eta}{\textstyle \kappa}\,\dot{\bf w},
\end{array}
\right.
\label{LCBiot}
\end{equation}
where ${\bf u}=(u_x,\,u_y)^t$ is the solid displacement, ${\bf U}=(U_x,\,U_y)^t$ is the fluid displacement, ${\bf w}=\phi\,({\bf U}-{\bf u})=(w_x,\,w_y)^t$ is the relative displacement, ${\bf I}$ is the identity tensor, ${\bf \sigma}$ is the stress tensor, ${\bf \varepsilon}=\frac{1}{2}\,\left({\bf \nabla\,u}+{\bf \nabla}^t\,{\bf u}\right)$ is the strain tensor, and $p$ is the pore pressure. 


\subsection{Boundary conditions}\label{SecPSjc}

The governing equations (\ref{LCAcous}) and (\ref{LCBiot}) have to be completed by a set of boundary conditions along the $N$ plane interfaces $y=\alpha_j$ ($j=0,\,\cdots,N-1$). For this purpose, we take $[g]_j$ to denote the jump in a function $g$ from $\Omega_j$ to $\Omega_{j+1}$ across $\alpha_j$
\begin{equation}
\begin{array}{lll}
[g]_j&=&
\displaystyle
\lim_{\varepsilon\rightarrow 0,\varepsilon > 0}g(x,\,\alpha_j+\varepsilon,\,t)-\lim_{\varepsilon\rightarrow 0,\varepsilon>0}g(x,\,\alpha_j-\varepsilon,\,t),\\
[8pt]
&=& (g)_j^+-(g)_{j}^-.
\end{array}
\label{NotationBC}
\end{equation}
The fluid / porous interface $\alpha_0$ is modeled with the following boundary conditions \cite{ROSENBAUM74,GUREVICH99,BOURBIE87,FENG83a,SHARMA08}
\begin{subnumcases}{\label{JC-FB}}
\displaystyle
(u_y)^-_0+(w_y)^-_0=(U_y)^+_0,\label{JC-FB1}\\
[8pt]
\displaystyle
(\sigma_{xy})^-_0=0,\label{JC-FB2}\\
[8pt]
\displaystyle
(\sigma_{yy})^-_0=-(p)_0^+,\label{JC-FB3}\\
[8pt]
\displaystyle
[p]_0=\frac{\textstyle 1}{\textstyle {\mathcal K}}\,(\dot{w}_y)^-_0.\label{JC-FB4}
\end{subnumcases}
Eq. (\ref{JC-FB4}) involves the hydraulic permeability of the interface $\mathcal{K}$, resulting in the following limit cases:
\begin{itemize}
\item if $\mathcal{K} \rightarrow +\infty$, then Eq. (\ref{JC-FB4}) becomes $[p]_0=0$, describing {\it open pores};
\item if $\mathcal{K}\rightarrow 0$, then no fluid exchange occurs across $\alpha_0$, and Eq. (\ref{JC-FB4}) is replaced by $(\dot{w}_y)^-_0=0$, describing {\it sealed pores};  
\item if $0<\mathcal{K}<+\infty$, then an intermediate state between open and sealed pores is reached, describing {\it imperfect pores}.
\end{itemize}
The porous / porous interfaces $\alpha_j$ ($j=1,\cdots N-1$) are assumed to be in perfect bonded contact \cite{GUREVICH99,BOURBIE87}:
\begin{equation}
[u_x]_j=0,\quad [u_y]_j=0,\quad [w_y]_j=0,\quad [\sigma_{xy}]_j=0,\quad [\sigma_{yy}]_j=0,\quad [p]_j=0.
\label{JC-BB}
\end{equation}


\section{Semi-analytical solution}\label{SecAnal}

\subsection{Helmholtz decompositions}\label{SecAnalFormu}

Pressure and stress components are eliminated from Eqs. (\ref{LCBiot}), giving a $(\textbf{u},\,\textbf{w})$ second-order wave formulation
\begin{equation}
\left\{
\begin{array}{l}
\displaystyle
(\lambda_0+\mu+m\,\beta^2)\,\nabla(\nabla.{\bf u})+\mu\,{\bf \nabla}^2\,{\bf u}+m\,\beta\,{\bf \nabla}({\bf \nabla}.{\bf w})=\rho\,\ddot{\bf u}
+\rho_f\,\ddot{\bf w},\\
[10pt]
\displaystyle
m\,\beta\,{\bf \nabla}({\bf \nabla}.{\bf u})+m\,{\bf \nabla}({\bf \nabla}.{\bf w})=\rho_f\,\ddot{{\bf u}}+\frac{\textstyle a\,\rho_f}{\textstyle \phi} \ddot{{\bf w}}+\frac{\textstyle \eta}{\textstyle \kappa}\,\dot{{\bf w}}.
\end{array}
\right. 
\label{AlembertBiot}
\end{equation}
The solid and relative displacements in Eqs. (\ref{AlembertBiot}) can be expressed by
\begin{equation}
{\bf u}={\bf \nabla}\,\varphi+{\bf \nabla}\,\times{\bf \Psi},\qquad {\bf w}={\bf \nabla}\,\varphi^r+{\bf \nabla}\,\times{\bf \Psi}^r,
\label{Potentiels}
\end{equation}
where $\varphi$ and $\varphi^r$ are scalar potentials, and ${\bf \Psi}$ and ${\bf \Psi}^r$ are vector potentials. One introduces mass, stiffness and damping matrices
\begin{equation}
\begin{array}{l}
{\bf K_P}=\left(
\begin{array}{cc} 
\lambda_0 + 2\mu+ m\,\beta^2 & m\,\beta \\
[6pt]
m\,\beta & m 
\end{array} 
\right),\quad
{\bf K_S}= 
\left( 
\begin{array}{cc} 
\mu & 0 \\
[6pt]
0 & 0 
\end{array} 
\right),\\
\\
{\bf M}=\left( 
\begin{array}{cc} 
\rho & \rho_f \\
[6pt]
\rho_f & \displaystyle \frac{\textstyle a_\infty\,\rho_f}{\textstyle \phi} 
\end{array} 
\right),\quad
{\bf C}=\left( 
\begin{array}{cc} 
0 & 0 \\
[6pt]
0 &  \displaystyle \frac{\textstyle \eta}{\textstyle \kappa}
\end{array} 
\right).
\end{array}
\label{Matrices}
\end{equation}
The Fourier transforms in time $t$ and space $x$ are defined by
\begin{equation}
\begin{array}{l}
\displaystyle
f(t)=\int_{-\infty}^{+\infty}f^*(\omega)\,\exp(-i \omega t)d\omega,\hspace{1cm}
f^*(\omega)=\frac{1}{2 \pi}\int_{-\infty}^{+\infty}f(t)\,\exp(i \omega t)dt,\\
[10pt] 
\displaystyle
g(x)=\int_{-\infty}^{+\infty}\overline{g}(k_x)\,\exp(i k_x x)\,dk_x, \hspace{1.1cm}
\overline{g}(k_x)=\frac{1}{2 \pi}\int_{-\infty}^{+\infty}\,g(x)\exp(-i k_x x)dx.
\label{Fourier}
\end{array}
\end{equation}
Applying these $x$ and $t$ Fourier transforms  to relations (\ref{AlembertBiot}) and (\ref{Potentiels}) yields decoupled ordinary differential systems in the frequency-wavenumber domain, associated with fast and slow compressional waves $P_f$ and $P_s$, and with shear wave $S$, respectively
\begin{equation}
\begin{array}{l}
\displaystyle
\left(-\left(\frac{d^2}{dy^2}-k_x^2\right)\,{\bf K_P}
-\omega^2\,{\bf M}- i\,\omega\,{\bf C}\right) 
\left(
\begin{array}{c}
\overline\varphi^* \\ 
[6pt]
\overline\varphi^{r*}
\end{array}
\right)=\left(
\begin{array}{c} 
0 \\
[6pt]
0 
\end{array}
\right),\\
\\
\displaystyle
\left(-\left(\frac{d^2}{dy^2}-k_x^2\right)\,{\bf K_S}
-\omega^2\,{\bf M}-i\,\omega\,{\bf C}
\right)\left(
\begin{array}{c} 
\overline{{\bf \Psi}}^* \\
[6pt]
\overline{{\bf \Psi}}^{r*}
\end{array}
\right)=\left(
\begin{array}{c} 
\textbf{0} \\
[6pt]
\textbf{0} 
\end{array}\right).
\end{array}
\label{ODE}
\end{equation}
Taking the Fourier transforms of Eqs. (\ref{ODE}) over $y$ gives the dispersion relations 
\begin{equation} 
\det \left(k_{Pj}^2\,{\bf K_P}-\omega^2{\bf M}-i\,\omega\,{\bf C} \right)=0,\qquad
k_S^2=\frac{\omega^2}{\mu} \left(\rho+\rho_f \,G\right),
\label{Dispersion}
\end{equation}
where $k_{Pj}$ ($j=f,s$) are the wavenumbers of the fast ($f$) and slow ($s$) compressional waves, and $k_{S}$ is the wavenumber of the shear wave, see \ref{SecAppendixMatrices}, Eq. (\ref{eq:wavenbs}). $G$ is defined in  \ref{SecAppendixMatrices}, Eq. (\ref{FG}). This gives the phase velocities $c_{Pj}=\omega\,/\,\Re(k_{Pj})$ and $c_S=\omega\,/\,\Re(k_{S})$, where $c_{Pf}>\max(c_{Ps},\,c_S)$.


\subsection{Exact stiffness matrix approach}\label{SecAnalStiff}

The exact stiffness matrix approach is based on vectors of transformed displacement and stress components \cite{DEGRANDE98}, defined as 
$$
\overline{{\bf \Upsilon}}^*=(\overline{u}^*_x,\,i\,\overline{u}^*_y,\,i\,\overline{w}^*_y)^t,\qquad
\overline{{\bf \Sigma}}^*=(\overline{\sigma}^*_{xy}, \,i\,\overline{\sigma}^*_{yy}, \,
-i\,\overline{p}^* )^t.
$$
Analytical expression for the transformed displacement is then  solution of the matrix system
\begin{equation}
\underbrace{
\left(
\begin{array}{cc} 
 {\bf S}^T                              & \hspace{-0.3cm}{\bf S}^R {\bf Z}(h_j) \\
[4pt]
\hspace{-0.3cm}-{\bf S}^T {\bf Z}(h_j)  & \hspace{-0.3cm}-{\bf S}^R 
\end{array}
\hspace{-0.2cm}\right)
\hspace{-0.1cm}
\left(
\begin{array}{cc}
\hspace{-0.3cm}{\bf Q}^T                & \hspace{-0.3cm}{\bf Q}^R {\bf Z}(h_j) \\ 
[4pt]
\hspace{-0.2cm}{\bf Q}^T {\bf Z}(h_i)   & \hspace{-0.1cm}{\bf Q}^R
\end{array}
\hspace{-0.2cm}\right)^{-1}}_{{\bf T}^j_{(6 \times 6)}}
\hspace{-0.1cm}
\left(
\begin{array}{l}  
\hspace{-0.2cm}(\overline{{\bf \Upsilon}}^*)_{j-1}^-  \\
[4pt]
\hspace{-0.2cm}(\overline{{\bf \Upsilon}}^*)_j^+
\end{array}
\hspace{-0.2cm}\right)
=\left(
\begin{array}{c}  
 (\overline{{\bf \Sigma}}^*)_{j-1}^- \\
 [4pt]
\hspace{-0.4cm}-(\overline{{\bf \Sigma}}^*)_j^+
\end{array}
\hspace{-0.2cm}\right), 
\label{parite}
\end{equation}
where $h_j=\alpha_{j-1}-\alpha_j$ is the thickness of layer $\Omega_j$ ($j=1,\,\cdots,N-1$). Further information about the matrices ${\bf Z}(h_j)$, ${\bf S}^{T,R}$, ${\bf Q}^{T,R}$ is given in \ref{SecAppendixMatrices}, Eqs. (\ref{MatZ}), (\ref{MatS}) and (\ref{MatQ}), respectively. The superscripts $T$ and $R$ stand for transmitted and reflected waves respectively. Special attention is paid to the conditioning of the matrices. Increasing exponential terms corresponding to reflected waves are excluded from  the formulation. These are included in the unknown wave potential vectors, which do not need to be calculated, see \cite{MESGOUEZ09}. Only decreasing exponential terms are therefore explicitly included in the expressions of ${\bf Z}(h_j)$, \ref{SecAppendixMatrices}, Eq. (\ref{MatZ}). Based on Eqs. (\ref{JC-BB}), a classical assembling procedure between the porous layers is used, which involves the continuity of the stresses and displacements at each interface. In the case of the half-space (\textit{hs}) $\Omega_N$ under the layers, only the transmitted terms are kept and ${\bf T}^{hs}={\bf S}^T({\bf Q}^T)^{-1}$. The whole matrix system involving all the poroelastic subdomains has dimension $3\,N \times 3\,N$. 


\subsection{Assembling of the fluid / porous interface $\alpha_0$}\label{SecAssFlu}

In the fluid domain $\Omega_0$, the acoustic pressure $\overline{p}^*$ is the sum of an incident cylindrical wave emitted by the source point and a reflected wave
\begin{equation}
\overline{p}^*=\overline{p}^{*I}+\overline{p}^{*R}=\frac{\textstyle i\,S(\omega)}{\textstyle 4\,\pi}\,\frac{\textstyle \exp(i\,k_{yf}\,\mid y-y_{s} \mid)}{\textstyle k_{yf}}+ A\,\exp(i\,k_{yf}\,y),
\label{eq:presfluide0}
\end{equation}
where $A$ is determined by the boundary conditions (\ref{JC-FB}) along $\alpha_0$, and $k_f^2=\omega^2/c_f^2=k_x^2+k^2_{yf}$. Taking $k_{yf}$ with $\Im (k_{yf})\geqslant 0$, the pressure vanishes at $y\rightarrow+\infty$ (the radiation condition). Expressions for $\overline{U}^*_x$ and $\overline{U}^*_y$ are easily deduced from Eqs. (\ref{eq:presfluide0}) and (\ref{LCAcous}). In the matrix condensed form, the part to be included into the overall system is written as follows
\begin{equation}
\left( 
\begin{array}{cc}
\displaystyle
-\frac{\textstyle i\,\rho_f\,\omega^2}{\textstyle k_{yf}} &
\displaystyle 
-\frac{\textstyle i\,\rho_f\,\omega^2}{\textstyle k_{yf}}\\
[10pt]
\displaystyle
-\frac{\textstyle i\,\rho_f\,\omega^2}{\textstyle k_{yf}} & 
\displaystyle
-\frac{\textstyle i\,\rho_f\,\omega^2}{\textstyle k_{yf}}
-\frac{\textstyle i\,\omega}{\textstyle \mathcal{K}}    
\end{array}
\right)
\left(
\begin{array}{c}
\displaystyle
i\,(\overline{u}^*)_0^- \\ 
[10pt]
\displaystyle
i\,(\overline{w}^*)_0^-
\end{array}\!\!
\right)=
\left(
\begin{array}{c}
\displaystyle
i\,(\overline{p}^*)_0^+ +\frac{\textstyle S(\omega)}{\textstyle 2\,\pi}\,\frac{\textstyle \exp(i\,k_{yf}\,y_s)}{\textstyle \,k_{yf}} \\
[10pt]
\displaystyle
i\,(\overline{p}^*)_0^- +\frac{\textstyle S(\omega)}{\textstyle 2\,\pi}\,\frac{\textstyle \exp(i\,k_{yf}\,y_s)}{\textstyle k_{yf}}
\end{array}
\right).
\label{SystemFluide}
\end{equation}
Coupling system (\ref{parite}) with (\ref{SystemFluide}) yields the overall system 
\begin{equation}
\left(
\begin{array}{cccc}
...\!\!& ...     & ...   & ...   \\
[4pt]
...\!\!& 
\displaystyle T^1_{22} -\frac{\textstyle i\,\rho_f\,\textstyle \omega^2}{\textstyle k_{yf}} & \displaystyle T^1_{23} -\frac{\textstyle i\,\rho_f\,\omega^2}{\textstyle k_{yf}}& ...\\
[10pt]
...\!\!& 
\displaystyle T^1_{32} -\frac{\textstyle i\,\rho_f\,\omega^2}{\textstyle k_{yf}} & 
\displaystyle T^1_{33} -\frac{\textstyle i\,\rho_f\,\omega^2}{\textstyle k_{yf}} - \frac{\textstyle i\,\omega}{\textstyle \mathcal{K}}&... \\
[4pt]
\!\!...&... &... &... \\
[4pt]
\!\!...& ...     & ...   & ...   
\end{array}
\!\!\right)\!\!
\left(
\!\!\!\! 
\begin{array}{c}
   (\overline{u}^*_x)_0^-\\ 
[8pt]
i\,(\overline{u}^*)_0^-\\
[8pt]
i\,(\overline{w}^*_{y})_0^-\\
[4pt]
...\\
[4pt]
...
\end{array}
\!\!\!\!
\right)
\!\!=\!\!
\left(
\!\!\!\!
\begin{array}{c}
0\\
[4pt]
\displaystyle
\frac{\textstyle S(\textstyle \omega)}{\textstyle 2\,\pi}\,\frac{\textstyle \exp(i\,k_{yf}\,y_s)}{\textstyle \,k_{yf}}\\
[10pt]
\displaystyle
\frac{\textstyle S(\textstyle \omega)}{\textstyle 2\,\pi}\,\frac{\textstyle \exp(i\,k_{yf}\,y_s)}{\textstyle \,k_{yf}}\\
[4pt]
0\\...
\end{array}
\!\!\!\!
\right).
\label{eq:globalsystem}
\end{equation}
The solution to system (\ref{eq:globalsystem}) gives the horizontal, vertical solid and relative displacements at each interface. The displacements, stresses, velocities and acoustic pressure in each domain $\Omega_j$ can then be obtained.


\subsection{Processing the oscillating integrals}\label{SecIntegrals}

Carrying out inverse Fourier transforms over the horizontal wavenumber $k_x$ and the angular frequency $\omega$ requires the use of numerical integration procedures.

\begin{figure}[htbp]
\begin{center}
\begin{tabular}{cc}
\includegraphics[scale=0.38]{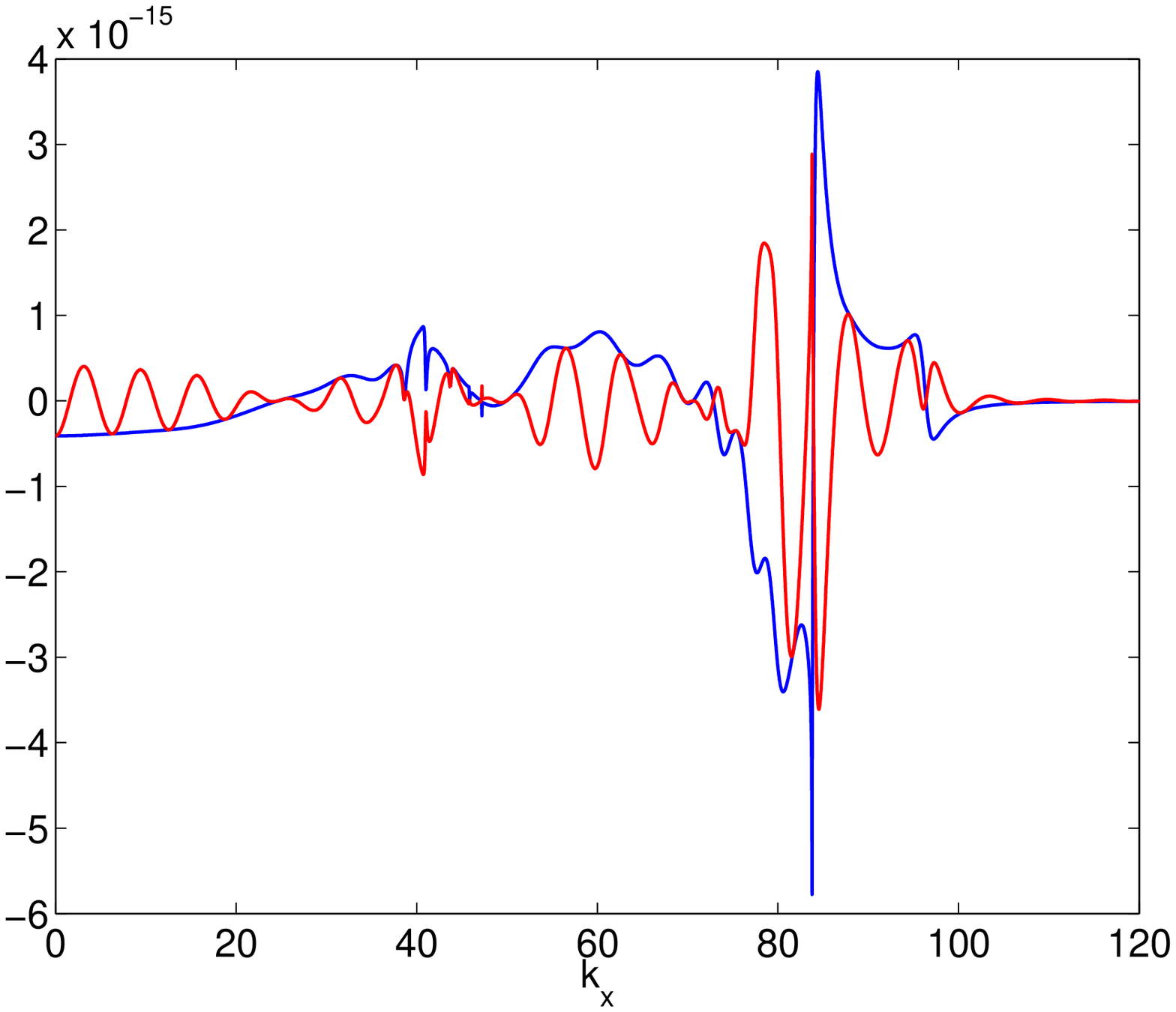} &
\includegraphics[scale=0.38]{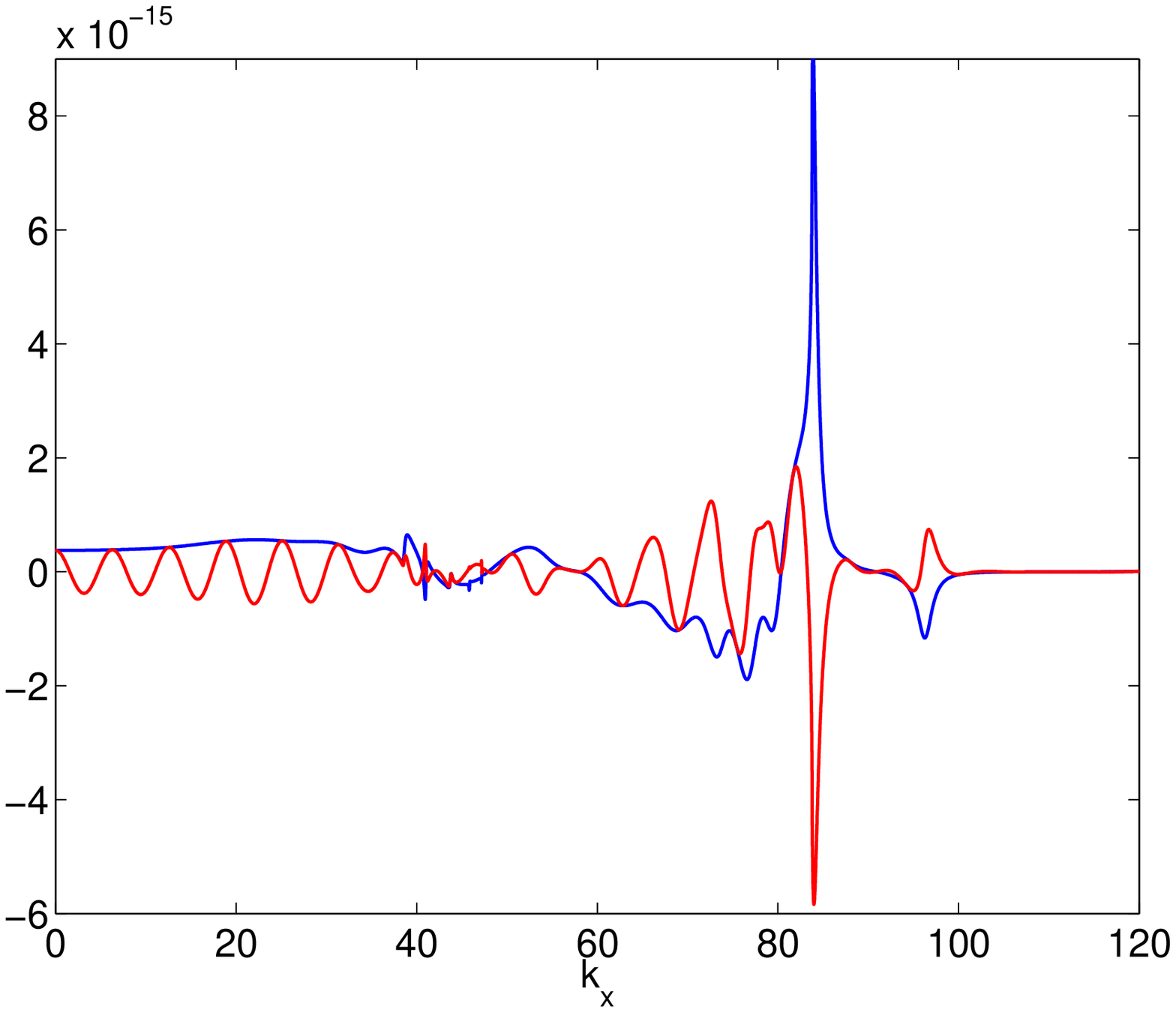} 
\end{tabular}
\end{center}
\vspace{-.7cm}
\caption{Processing the integrals. Real part (left) and imaginary part (right) of $\overline{\dot{u}_y}^*(k_x,\,\omega_0)$ (blue) and 
$\overline{\dot{u}_y}^*(k_x,\,\omega_0)\,\cos(k_x\,x)$ (red). This signal is obtained at recorder $R_4$ in the case of $N=4$ interfaces, with sealed pore conditions; see section \ref{SecRes} for details.}
\label{figIntegrande}
\end{figure}

Two problems have to be overcome when calculating the integral over the horizontal wavenumber. First, the integrand shows fast oscillatory behavior due to the factor $\exp(i k_x x)$. Secondly, the envelope of the maximum amplitudes shows sharp peaks, as shown in Fig. \ref{figIntegrande}. Alternative techniques to the usual Fast Fourier Transform for evaluating the integral properly can be found in \cite{BOUCHON77,APSEL83,DRAVINSKI88} in the context of elastodynamics, and \cite{RAJAPAKSE95,DEGRANDE98} in that of poroelastodynamics, for example. Here we use the adaptive Filon  quadrature, which is particularly accurate and suitable for dealing with these integrals. The transformed fluid pressure, pore pressure, stress tensor normal component, vertical displacement and velocity are symmetric terms with respect to $k_x$. These functions, which are denoted by $\overline{h}^*$, can be written as follows
\begin{equation}
h(x,\,y,\,t)=\int_{-\infty}^{+\infty}\left\{2\int_{0}^{+\infty}\overline{h}^*(k_x,\,y,\,\omega) \cos(k_x\,x)\,dk_x \right \}\exp(-i\,\omega\,t)\,d\omega. 
\label{}
\end{equation}
In line with the Filon quadrature theory, the integrand is separated into the product of a slowly-varying component and a fast-oscillating one. The oscillatory kernel is rigorously accounted for, since only the complex function  $\overline{h}^*(k_x,\,y,\,\omega)$ is replaced by a (second degree here) polynomial approximation. The resulting integral is then computed analytically \cite{CHASE69}. 

One possible adaptive procedure consists in dividing the entire interval into several parts based on what is known about $\overline{h}^*(k_x,\,y,\,\omega)$. Because of the sharp changes in the integrand occuring around the wavenumbers of the propagating waves, the wavenumbers are calculated and sorted out. The quadrature is performed by discretizing finely in the neighborhood of these wavenumbers and more coarsely farther away. The integral is truncated depending on the highest wavenumber and adapted to each frequency.  

Concerning the numerical processing over $\omega$, the physical space-time domain $h(x,\,y,\,t)$ term is  real. We therefore obtain 
\begin{equation}
h(x,\,y,\,t)=\int_{0}^{+\infty}2\,\Re\left(\left\{2\int_{0}^{+\infty} \overline{h}^*(k_x,\,y,\,\omega)\,\cos(k_x \,x)\,dk_x \right\}\,\exp(-i\,\omega\,t)\right)\,d\omega. 
\end{equation}
Numerical integration is done using a Simpson quadrature, and the Nyquist-Shannon sampling theorem is checked. 


\section{Numerical solution}\label{SecNum}

\subsection{Numerical scheme}\label{SecNumScheme}

Velocity-stress formulation for the governing equations can be obtained from Eqs. (\ref{LCAcous}) and (\ref{LCBiot}). Setting
\begin{equation}
\bf{\Lambda}=
\left\{
\begin{array}{ll}
\displaystyle
\left(\dot{U}_x,\,\dot{U}_y,\,p\right)^t\qquad \hspace{3.7cm}\mbox{ in } \Omega_0,\\
[10pt]
\left(\dot{u}_x,\,\dot{u}_y,\,\dot{w}_x,\,\dot{w}_y,\,\sigma_{xx},\,\sigma_{xy},\,\sigma_{yy},\,p\right)^t\qquad \mbox{ in } \Omega_j,\quad j=1,\cdots,N,
\end{array}  
\right.
\label{VecU}
\end{equation}
gives the first-order linear system of partial differential equations with source term
\begin{equation}
\dot{\bf \Lambda}+{\bf A}\,\frac{\textstyle \partial}{\textstyle \partial x}\,{\bf \Lambda}+{\bf B}\,\frac{\textstyle \partial}{\textstyle \partial y}\,\bf{\Lambda}=-{\bf C}\,\bf{\Lambda}+{\cal \textbf{F}}.
\label{SystHyp}
\end{equation}
In Eq. (\ref{SystHyp}), ${\bf A}$, ${\bf B}$ and ${\bf C}$ are $3\times3$ matrices in $\Omega_0$, and $8\times8$ matrices in $\Omega_j$ ($j=1,\cdots,N$); the vector ${\cal \textbf{F}}$ accounts for the acoustic source in Eqs.  (\ref{LCAcous}). In $\Omega_0$, ${\bf C}={\bf 0}$.  Due to the non-zero diffusive term ${\bf C}$ present in the poroelastic media $\Omega_j$, an unsplit discretization scheme for Eq. (\ref{SystHyp}) would not be suitable, leading to a penalizing time step. A much more efficient method, which was used in \cite{CHIAVASSA11}, consists in using a Strang splitting procedure. The following partial differential equations are solved successively
\begin{subnumcases}
\displaystyle
\dot{\bf \Lambda}+{\bf A}\,\frac{\textstyle \partial}{\textstyle \partial\,x}\,{\bf \Lambda}+{\bf B}\,\frac{\textstyle \partial}{\textstyle \partial\,y}\,{\bf \Lambda}={\bf 0}, \label{Strang1}\\
\displaystyle
\dot{\bf{\Lambda}}=-{\bf C}\,\bf{\Lambda},\label{Strang2}\\
[8pt]
\displaystyle
\dot{\bf{\Lambda}}={\cal \textbf{F}}.\label{Strang3}
\label{Strang}
\end{subnumcases}
The propagative part (\ref{Strang1}) is solved using an ADER 4 scheme \cite{SCHWARTZKOPFF04}  on a uniform Cartesian grid, with spatial mesh sizes $\Delta x, \Delta y$ and a time step $\Delta t$. This explicit two-step finite-difference scheme is accurate to the fourth-order in both time and space, and satisfies the optimum CFL stability condition. The diffusive part (\ref{Strang2}) is solved exactly in the poroelastic layers $\Omega_j$, see Eq.  (18) in \cite{CHIAVASSA11}. This step induces no stability restrictions. The fluid $\Omega_0$ has no diffusive part, because there is no attenuation. Lastly, numerical integration is performed on $\Omega_0$ to account for the source term (\ref{Strang3}). 


\subsection{Mesh refinement}\label{SecNumAMR}

In the low-frequency range, the slow compressional wave is a diffusive non-propagating solution with very small wavelength $\lambda_{Ps}$. A very fine spatial mesh is therefore required. Since this wave does not propagate and is always located near the interfaces, space-time mesh refinement is a good strategy. For this purpose, a steady-state version of the algorithm presented in \cite{BERGER98} is developed. Each interface is inserted into a refined grid, where both the spatial meshes and the time step are divided by the refinement factor $q$. This procedure ensures that the same CFL number will be obtained in each grid. The coupling between coarse and fine meshes is achieved by performing spatial and temporal interpolations. The factor $q$ is used to obtain the same discretization of the slow wave on the refined zone, as with the fast wave on the coarse grid. Taking $f_0$ to denote the central frequency of the signal, direct calculations give $q(f_0)={c}_{Pf}(f_0)/c_{Ps}(f_0)$. 


\subsection{Immersed interface method}\label{SecNumIIM}

The discretization of the boundary conditions requires special care, for three reasons. First, if the interfaces do not coincide with the uniform meshing, then geometrical errors will occur. Secondly, the conditions (\ref{JC-FB}) and (\ref{JC-BB}) will not be enforced numerically by the finite-difference scheme, and the numerical solution will therefore not converge towards the exact solution. Lastly, the smoothness of the solution required to solve Eq. (\ref{SystHyp}) will not be satisfied across the interface, which will decrease the convergence rate of the ADER scheme.

To overcome these drawbacks without detracting the efficiency of Cartesian grid methods, an immersed interface method \cite{CHIAVASSA11,CHIAVASSA12} is used. The latter studies can be consulted for a detailed description of this method. The basic principle is as follows:  at the irregular nodes where the ADER scheme crosses an interface, modified values of the solution are used on the other side of the interface instead of the usual numerical values. 

Calculating these modified values is a complex task involving high-order derivation of boundary conditions (\ref{JC-FB})-(\ref{JC-BB}), Beltrami-Michell relations \cite{COUSSY95} and singular value decompositions. Fortunately, all these time consuming procedures can be carried out during a preprocessing stage and  only small matrix-vector multiplications need to be performed during the simulation. After optimizing the code, the extra CPU cost can be practically negligible, i.e. lower than 1\% of that required by the time-marching procedure. In addition, by choosing the order of the interface method carefully, it is possible to achieve the overall fourth-order accuracy of the ADER 4 scheme.


\section{Numerical results and discussion}\label{SecRes}

\subsection{Configuration and physical parameters}\label{SecResConf}

\begin{table}[htbp]
\begin{center}
\begin{tabular}{ll|l|l}
                 &                            & $\Omega_{2\,j-1}$   &  $\Omega_{2\,j}$      \\
\hline
Saturating fluid & $\rho_f$ (kg/m$^3$)        & 1000              &  1000               \\
                 & $c$ (m/s)                  & 1500              &  1500               \\
                 & $\eta$ (Pa.s)              & $10^{-3}$         &  $10^{-3}$          \\
                
Grain            & $\rho_s$ (kg/m$^3$)        & 2760              &  2644               \\
                 & $\mu$ (Pa)                 & $3.40\,10^9$      &  $7.04\,10^9$       \\
Matrix           & $\phi$                     & 0.24              &  0.20               \\ 
                 & $a_{\infty}$               & 2.3               &  2.4                \\
                 & $\kappa$ (m$^2$)           & $3.9\,10^{-13}$   &  $3.6\,10^{-13}$    \\
                 & $\lambda_0$ (Pa)           & $3.5\,10^9$       &  $5.5\,10^9$        \\
                 & $m$ (Pa)                   & $8.1\,10^9$       &  $9.7\,10^9$        \\
                 & $\beta$                    & 0.855             &  0.720               \\
\hline
Phase velocities & $c_{Pf}(f_0)$ (m/s)        & 2636.8            &  3263.7             \\
                 & $c_{Ps}(f_0)$ (m/s)        & 571.5             &  649.5              \\
				 & $c_S(f_0)$ (m/s)           & 1210.8            &  1751.0             \\
                 & $c_{St}$ (m/s)             & $1007.1$          &                     \\
                 & $f_c$ (kHz)                & 42.5              &  36.8               \\
\hline
\end{tabular}
\caption{Poroelastic media $\Omega_{2\,j-1}$ and $\Omega_{2\,j}$ $(j=1,\,\cdots,N)$: physical parameters and acoustic properties at $f_0=20$ kHz.}
\label{TabParametres}
\end{center}
\end{table}

Two configurations are proposed for testing the above semi-analytical and numerical methods. The influence of the various fluid / porous boundary conditions on the computed fields is also investigated. The first test involves a single interface ($N=1$) between a fluid and a porous medium. The second test is performed on a layered medium involving 4 interfaces ($N=4$). The accuracy and robustness of the semi-analytical and numerical approaches are proven in the case of realistic soil parameters.  

\begin{figure}[htbp]
\begin{center}
\begin{tabular}{cc}
\includegraphics[scale=0.35]{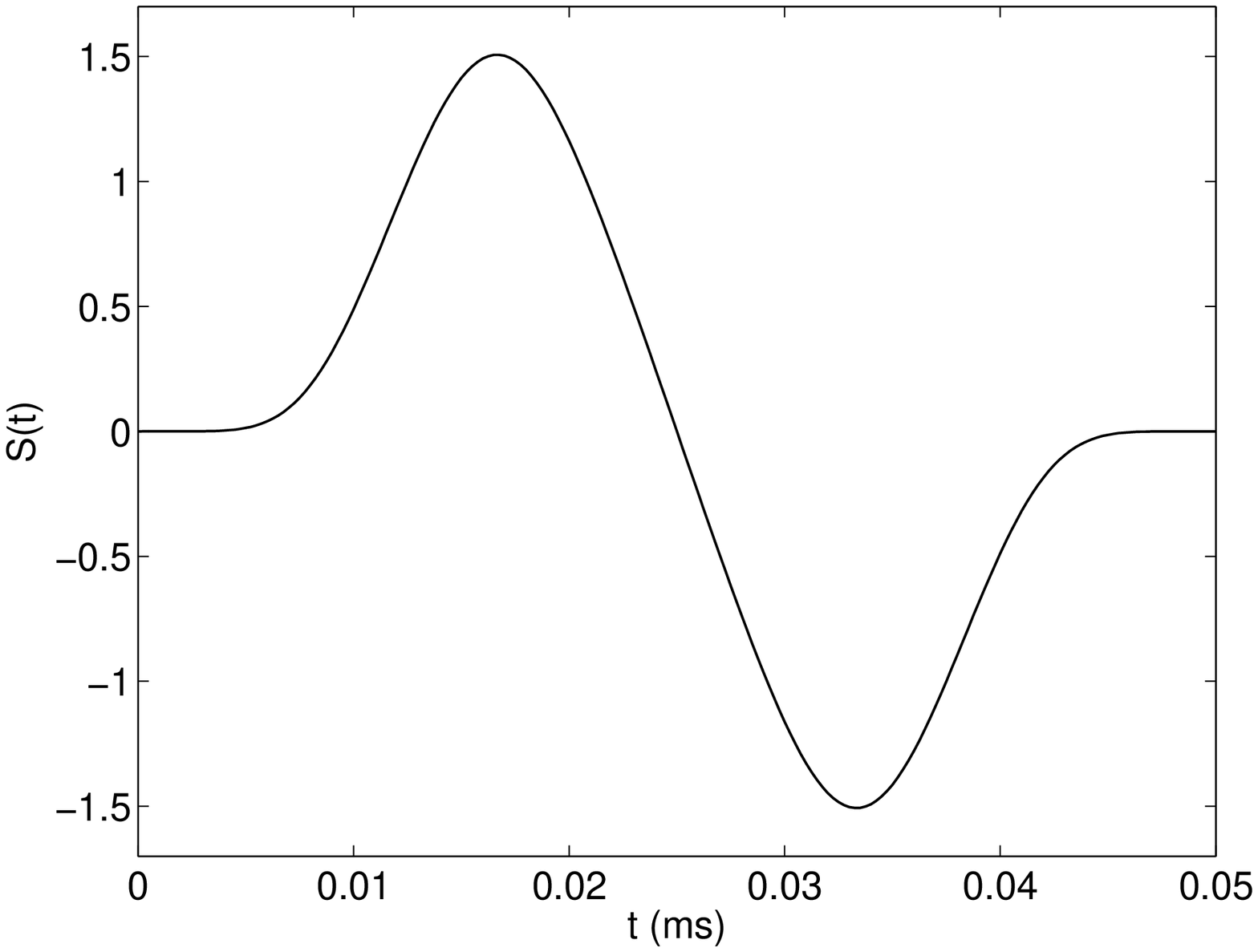} &
\includegraphics[scale=0.35]{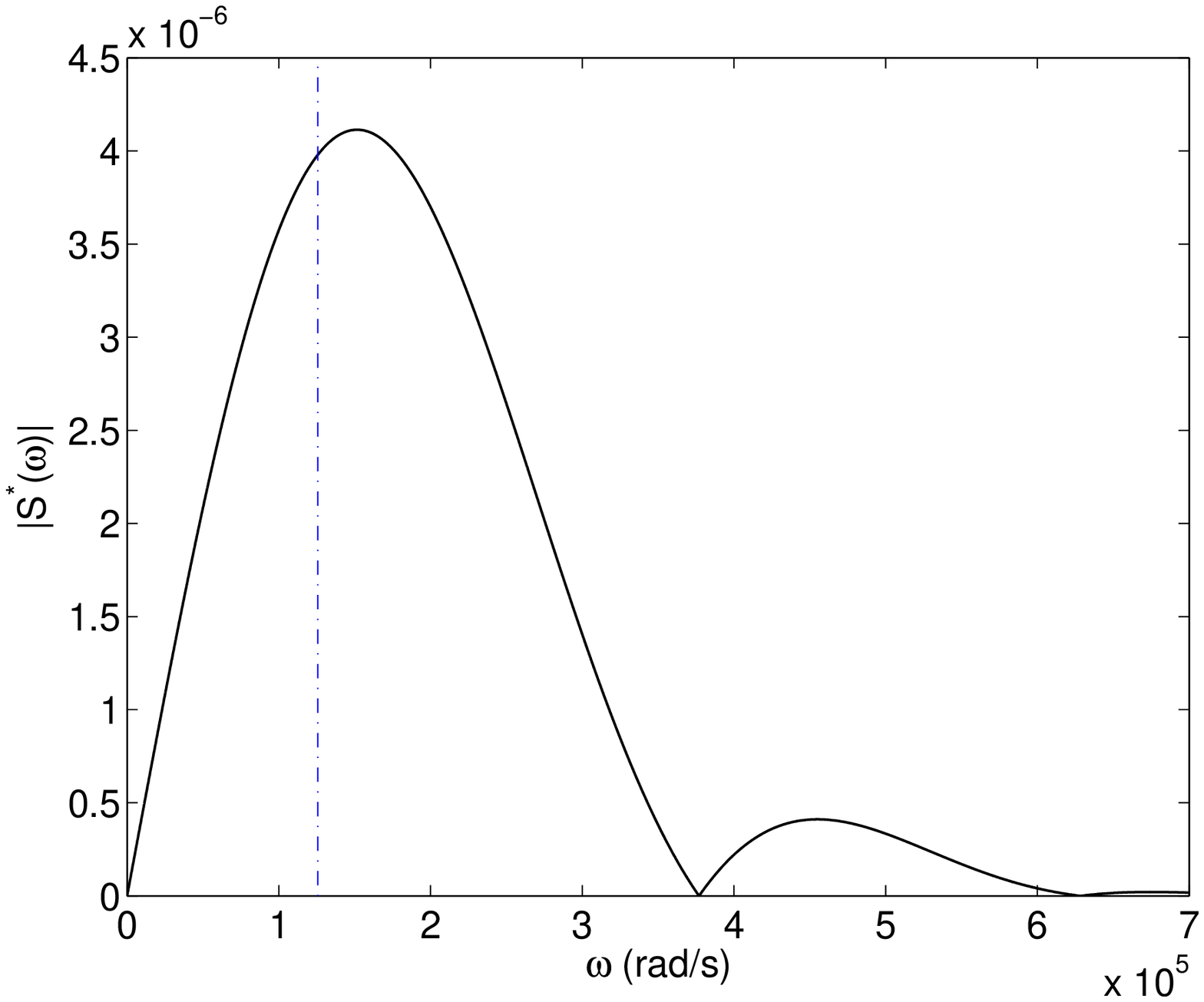} 
\end{tabular}
\end{center}
\vspace{-.5cm}
\caption{Time evolution (left) and spectrum (right) of the source $S$ defined by Eqs. (\ref{JKPS}) and (\ref{SpectreJKPS}). The vertical dashed line gives the central angular frequency $\omega_0=2\,\pi\,f_0$. }
\label{figSource}
\end{figure}

The acoustic medium $\Omega_0$ is water ($\rho_f=1000$ kg/m$^3$, $c=1500$ m/s). The poroelastic layers consist alternately of water-saturated sand in $\Omega_{2j-1}$ \cite{DENNEMAN02}, and Berea sandstone in $\Omega_{2j}$ \cite{MESGOUEZ09}. The poroelastic properties of these media are summarized in Table \ref{TabParametres}. In the case where an imperfect hydraulic contact occurs along the interface $\alpha_0$, the hydraulic permeability is $\mathcal{K}=5.\,10^{-8}$ m/s/Pa. 

The time evolution of the source $S$ appearing in Eq. (\ref{LCAcous}) is a combination of truncated sinusoids 
\begin{equation}
S(t)=
\left\{
\begin{array}{l}
\displaystyle
\displaystyle \sum_{m=1}^4 a_m\,\sin(\beta_m\,\omega_0\,t)\quad \mbox{ if  }\, 0<t<\frac{\textstyle 1}{\textstyle f_0},\\
[10pt]
0 \,\mbox{ otherwise}, 
\end{array}
\right.
\label{JKPS}
\end{equation}
where $\beta_m=2^{m-1}$ and $\omega_0=2\pi\,f_0$; the coefficients $a_m$ ensuring $C^6$ smoothness of the solution are: $a_1=1$, $a_2=-21/32$, $a_3=63/768$, $a_4=-1/512$. The Fourier transform of Eq. (\ref{JKPS}) is
\begin{equation}
S^*(\omega)=\sum_{m=1}^4 a_m\,\frac{\beta_m\,\omega_0}{2\,\pi}\,\frac{\exp(i\,2\,\pi\,{\omega_0}/{\omega})-1}{\omega^2-\beta_m\,\omega_0^2}.
\label{SpectreJKPS}
\end{equation}
The central frequency in Eqs. (\ref{JKPS}) and (\ref{SpectreJKPS}) is $f_0=20$ kHz, which is much smaller than the critical Biot frequency of poroelastic layers, see Eq. (\ref{Fc}) and Table \ref{TabParametres}.  The time evolution and Fourier transform of $S$ are presented in Fig. \ref{figSource}. In the following test cases, the source is always applied in the fluid domain at point $(x_s=0,\,y_s=4.10^{-3})$ m.


\subsection{Test 1: $N=1$ interface}\label{SecResN1}

\begin{figure}[htbp]
\begin{center}
\begin{tabular}{c}
\includegraphics[scale=0.60]{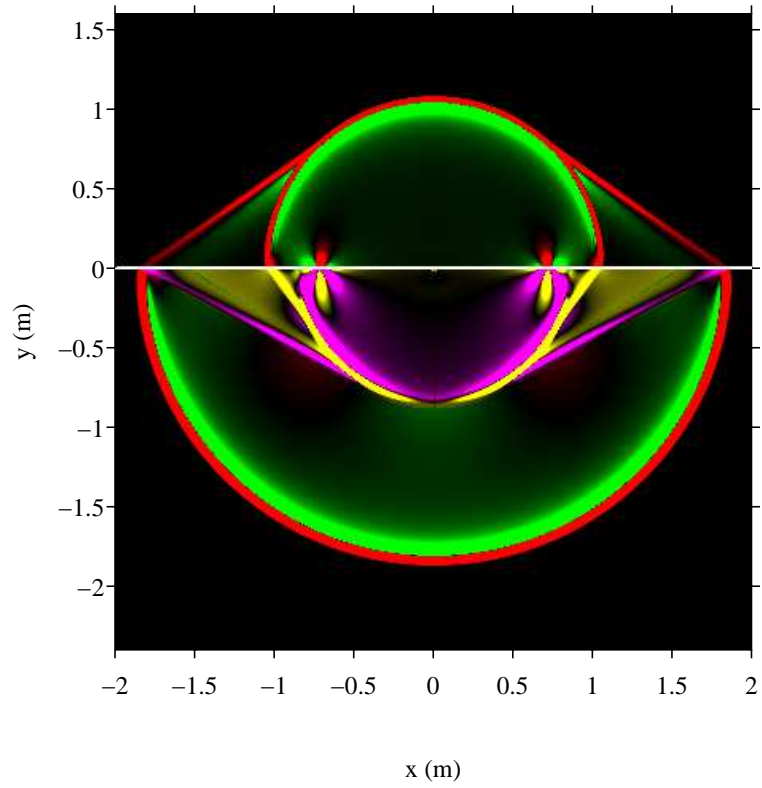} 
\end{tabular}
\end{center}
\vspace{-1cm}
\caption{$N=1$ interface with open pores. Snapshot of $p$ (in the fluid) and $\sigma_{xx}$ (in the poroelastic medium) at $t=0.72$ ms. Green-red: compressional waves; yellow-magenta: shear wave.}
\label{FigN1Carte}
\end{figure}

The interface between water and saturated sand is set at $\alpha_0=0$ m. With the present numerical method, the whole computational domain $[-2, 2]^2$ m$^2$ is discretized using a regular Cartesian grid with $\Delta x=\Delta y=2.10^{-3}$ m. This grid is refined locally  by a factor $q=5$ in the neighborhood of the interface in order to satisfy the criterion defined in section \ref{SecNumAMR}. With these parameters, the fast and slow compressional waves are discretized by at least 75 grid points per wavelength, and acoustic and shear waves by about 37 points per wavelength. 

\begin{figure}[htbp]
\begin{center}
\begin{tabular}{cc}
$p$ (open pores) & $\dot{u}_y$ (open pores)\\
\includegraphics[scale=0.35]{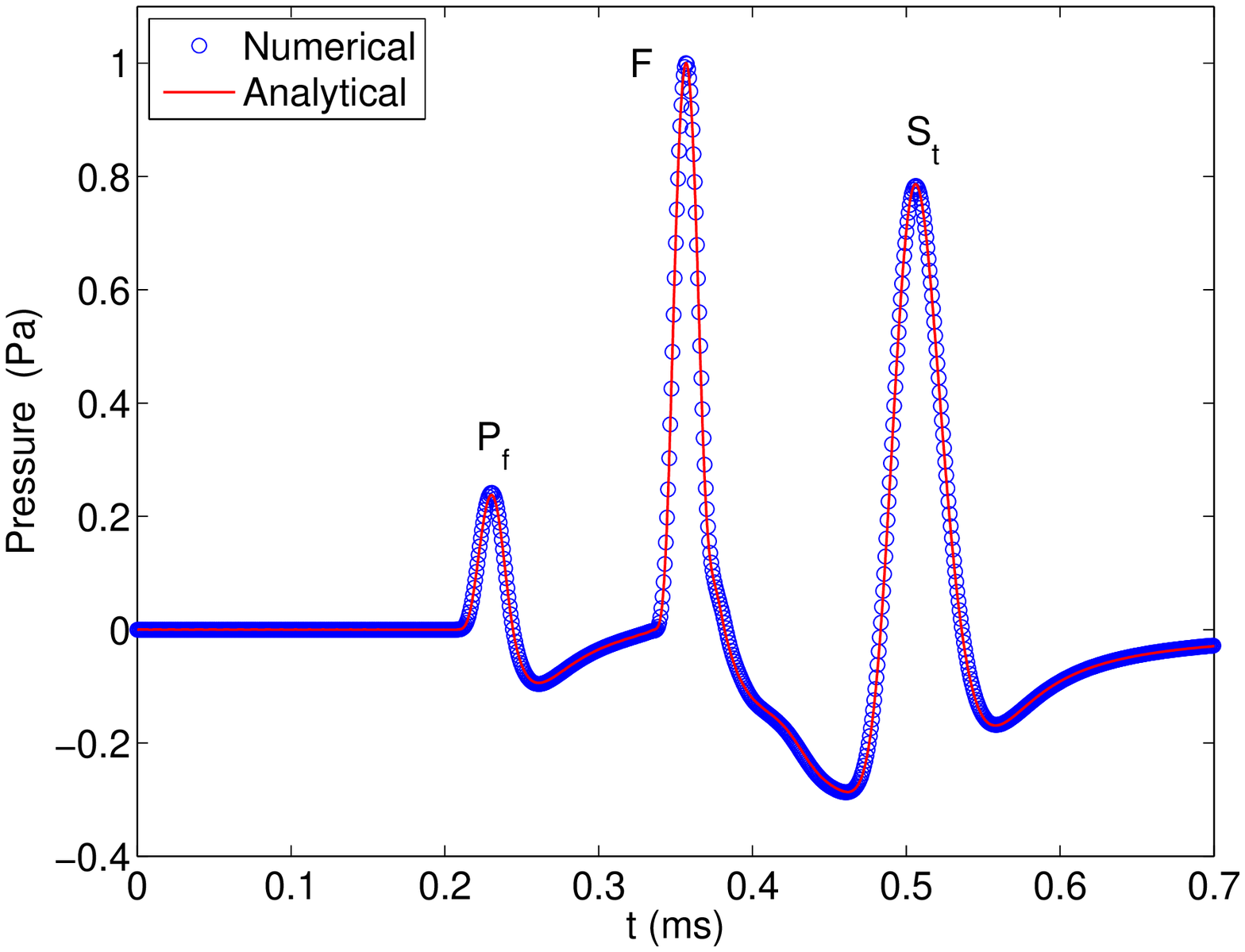} &
\includegraphics[scale=0.35]{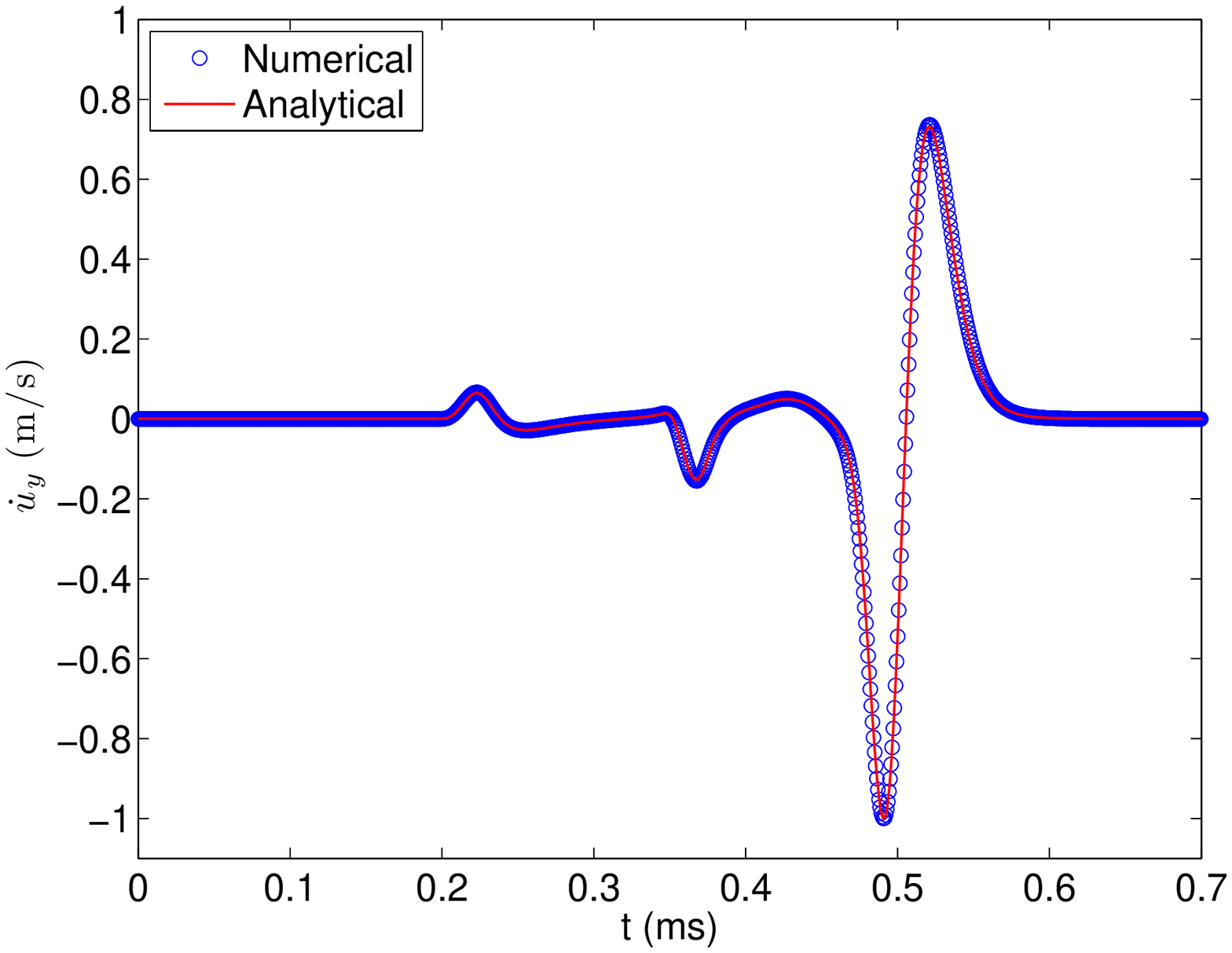}  \\ 
$p$ (imperfect pores) & $\dot{u}_y$ (imperfect pores)\\
\includegraphics[scale=0.35]{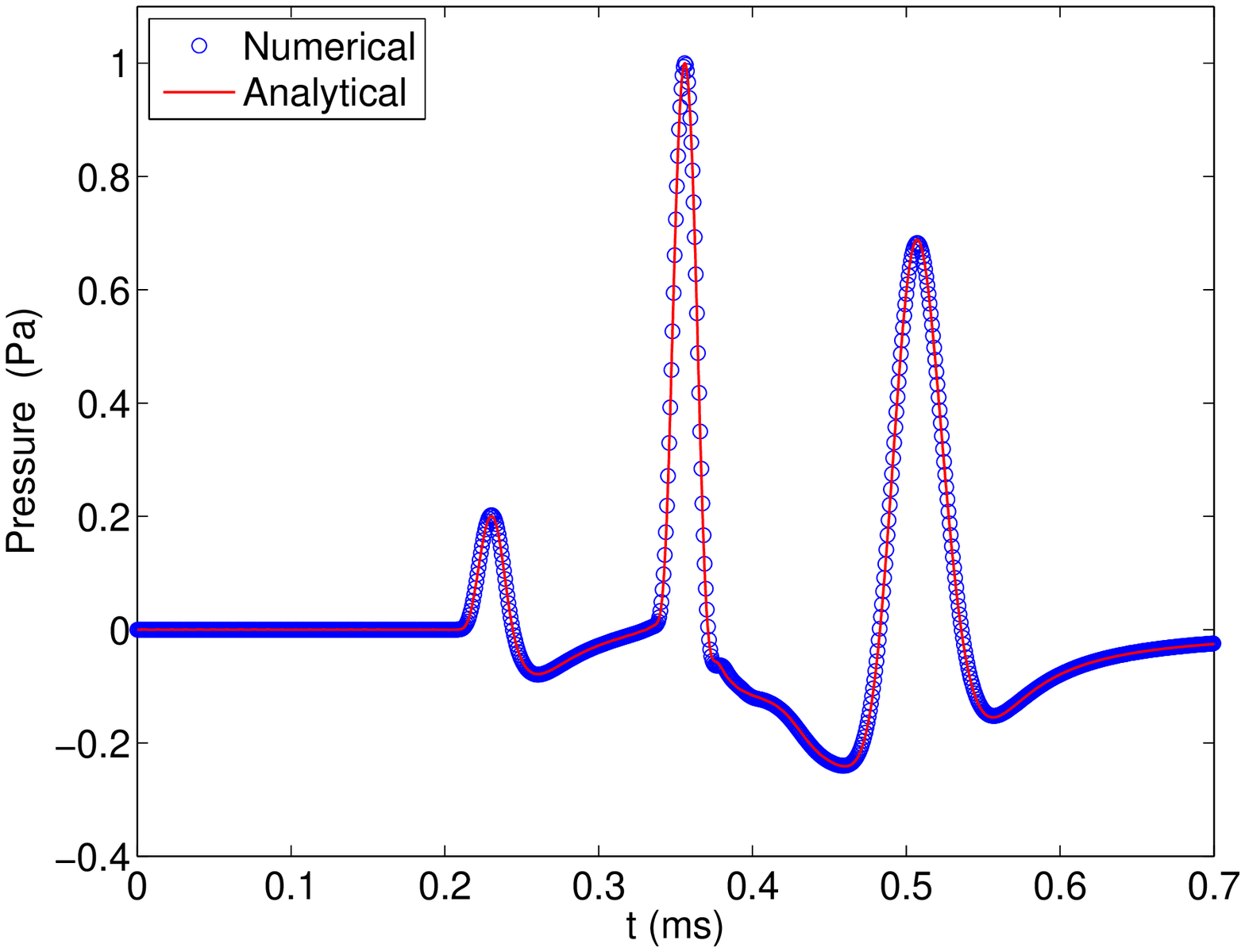}  &
\includegraphics[scale=0.35]{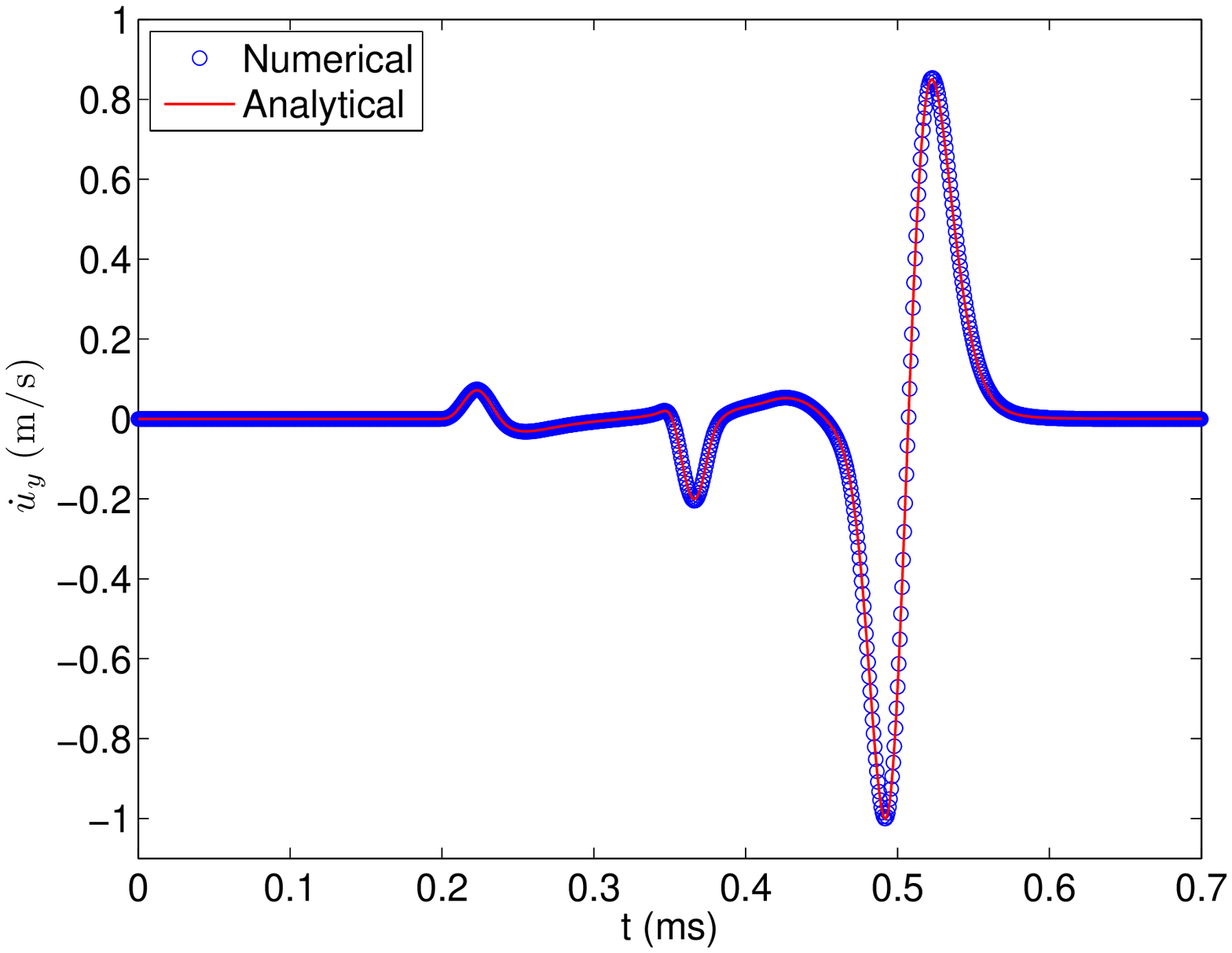}   \\ 
$p$ (sealed pores) & $\dot{u}_y$ (sealed pores)\\
\includegraphics[scale=0.35]{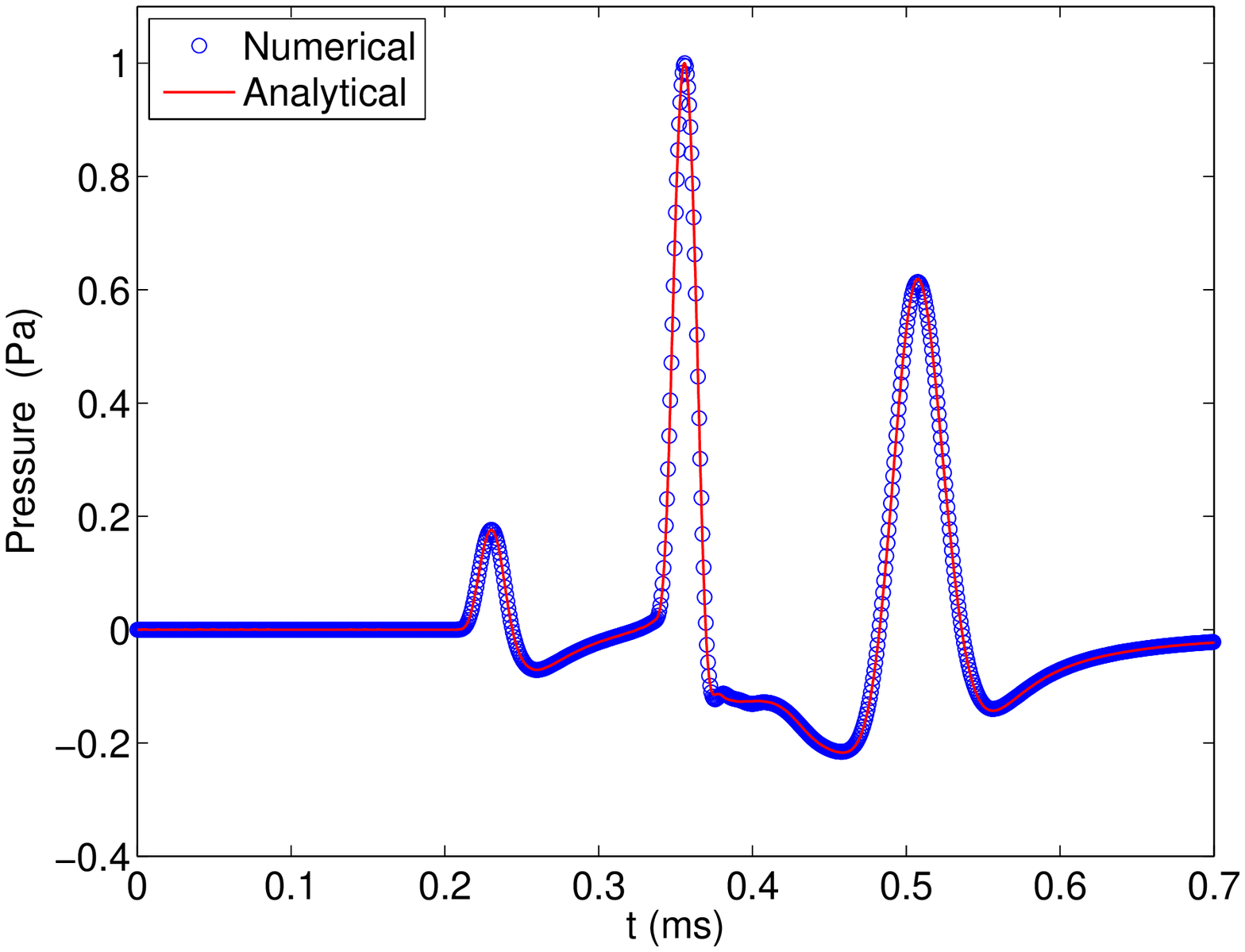} &
\includegraphics[scale=0.35]{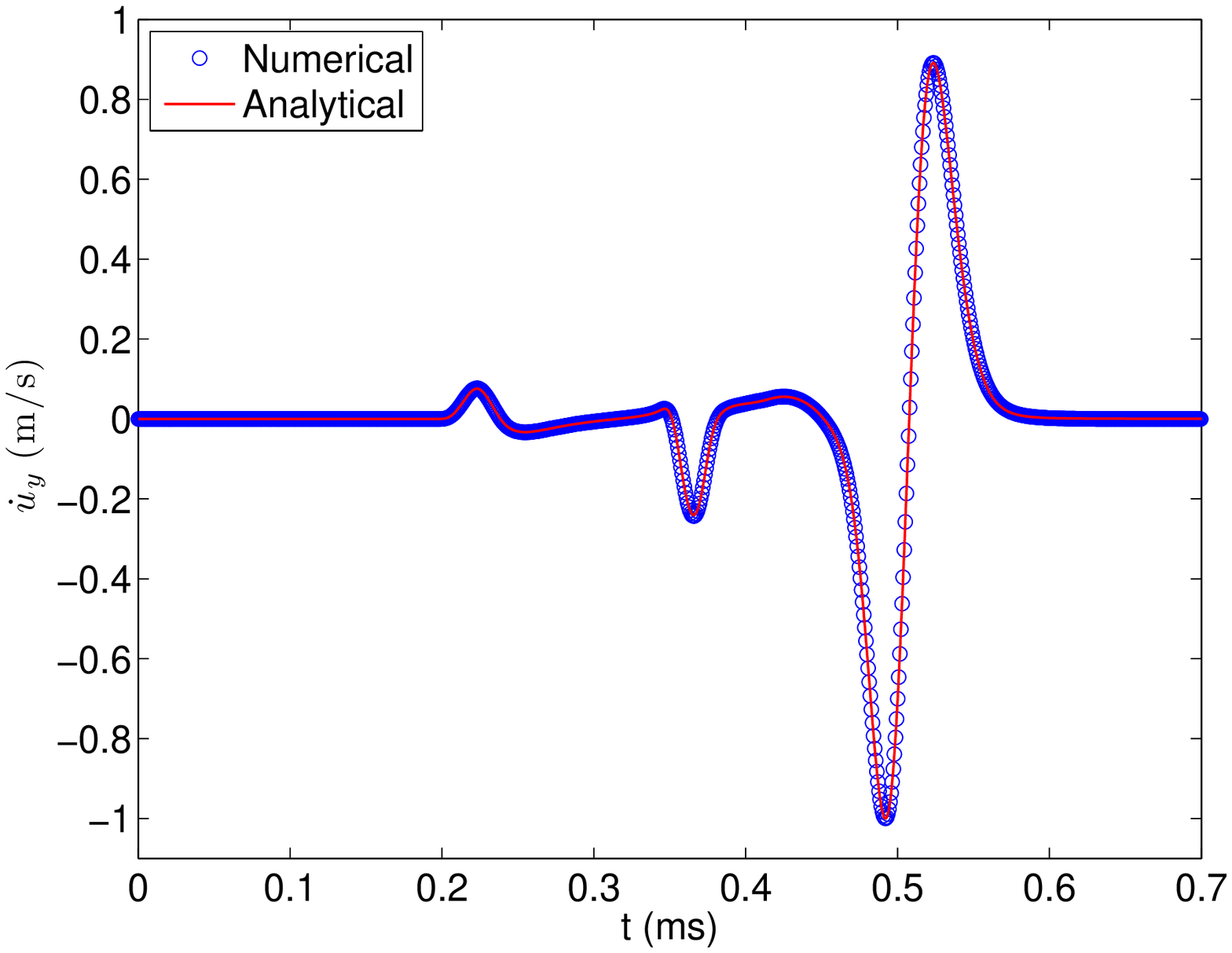}  \\ 
\end{tabular}
\end{center}
\vspace{-.5cm}
\caption{$N=1$ interface, with open pores (top), imperfect pores (middle) and sealed pores (bottom). Time evolution of the acoustic pressure $p$ at recorder R1 $(x=0.5, y=0.02)$ m (left) and that of the vertical velocity $\dot{u}_y$ at recorder R2 $(x=0.5, y=-0.012)$ m (right). }
\label{FigN1compare}
\end{figure}

The numerical solution obtained ($p$ in the fluid, $\sigma_{xx}$ in the poroelastic medium) is shown in Fig. \ref{FigN1Carte}, at time $t=0.72$ ms under open pore conditions. Since the source is located very near the interface, the reflected acoustic wave in $\Omega_0$ cannot be distinguished from the circular incident wave; the headwaves are also clearly visible. The transmitted fast compressional wave (circular wavefront) and shear wave (more complex structure) can be seen in $\Omega_1$. Pseudo-Stoneley waves can also be observed along the interface, just behind the shear wave. The transmitted slow compressional wave cannot be seen here; it remains located along the interface on very small spatial scales. 

In this configuration, the semi-analytical results were obtained with the following set of numerical parameters. As previously explained in section \ref{SecIntegrals}, wavenumbers corresponding to the propagating waves are calculated and sorted out. Let $k_{x,\min}$ and $k_{x,\max}$ denote the lowest and highest wavenumbers of propagating waves. The integral over $k_x$ is divided into three parts. The second part includes the sharp evolution of the integrand, and integration is performed in this part with a finer grid. We  perform 
\begin{enumerate}
\item integration over $k_x\in[0,\,0.75\,k_{x,\min}]$  with 500 subintervals; 
\item integration over $k_x\in[0.75\,k_{x,\min},\,1.25\,k_{x,\max}]$   with 5000 subintervals;
\item integration over $k_x\in[1.25\,k_{x,\max},\,20\,k_{x,\max}]$   with 200 subintervals. 
\end{enumerate}
The numerical integration over $\omega$ is led with 2000 regular subintervals, up to the maximum angular frequency $\omega_{\max}=7\,10^{5}$ rad.s$^{-1}$. Sampling with respect to time was performed with 4000 points, where $t\in[0,\,0.70]$ ms in Fig. \ref{FigN1compare}.   

\begin{figure}[htbp]
\begin{center}
\begin{tabular}{cc}
\includegraphics[scale=0.35]{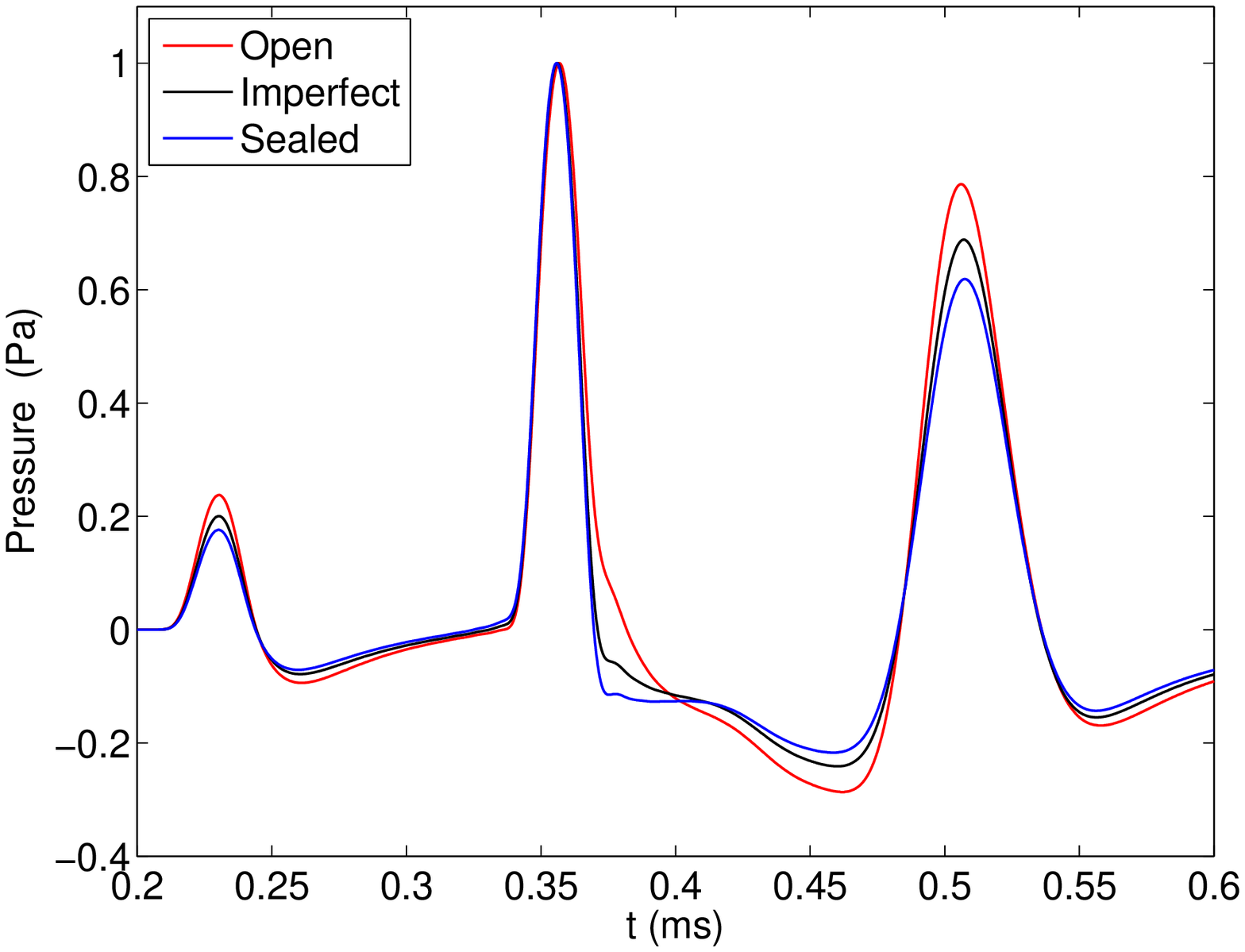} &
\includegraphics[scale=0.35]{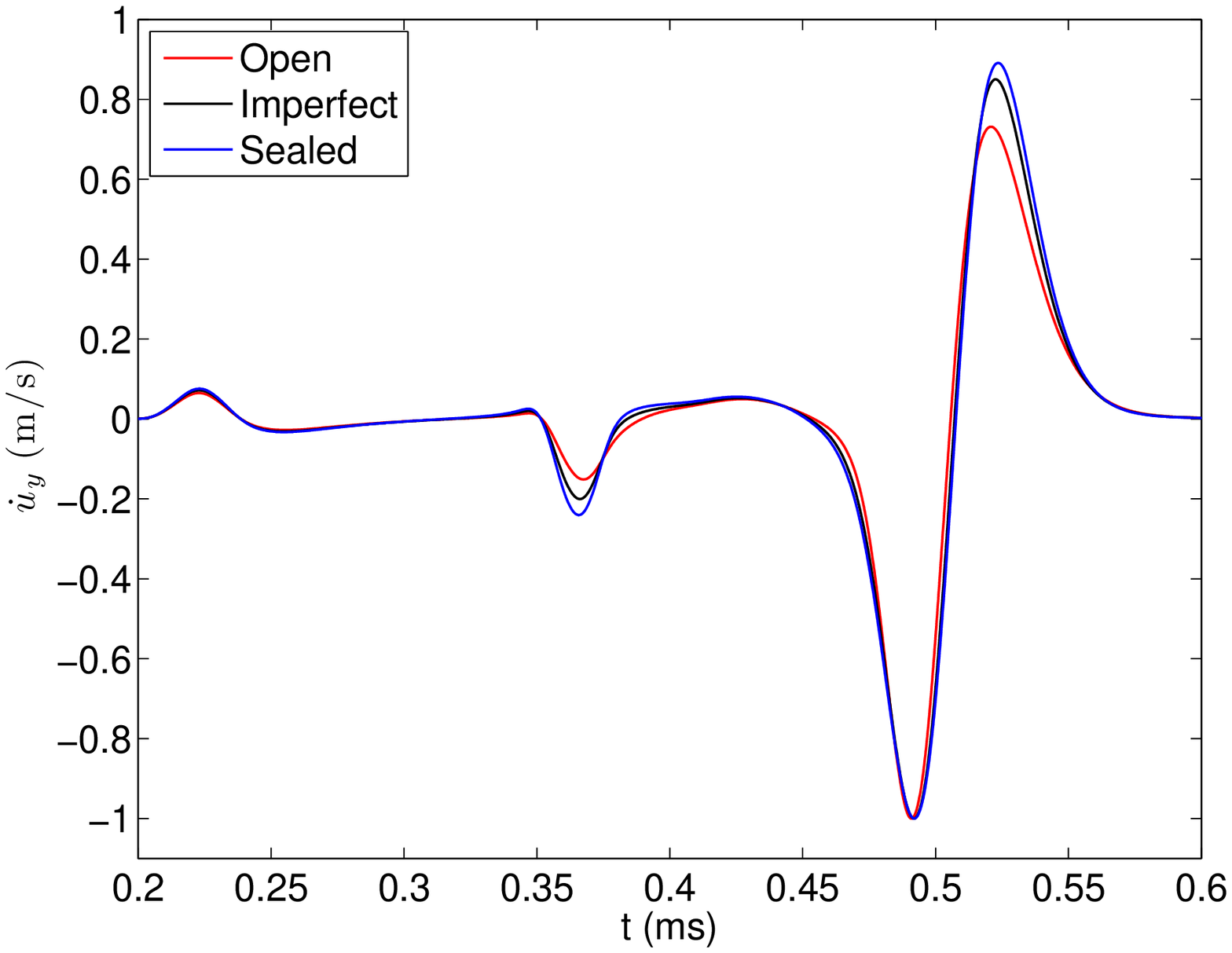} \\ 
\end{tabular}
\end{center}
\vspace{-.5cm}
\caption{$N=1$ interface. Comparison between the different pore  conditions: pressure (left) and vertical velocity (right). }
\label{FigN1pores}
\end{figure}

The time history of the pressure $p$  and of the vertical velocity $\dot{u}_y$ obtained with these two methods is presented in Fig. \ref{FigN1compare}. The receivers were placed near the interface in order to capture the surface waves. The pressure was recorded in the fluid domain at point R1 $(x=0.5, y=0.02)$ m while the vertical velocity was recorded in the porous medium at R2 $(x=0.5, y=-0.012)$ m.  Results were normalized with respect to the maximum value of the pressure (R1) or velocity (R2) respectively. Excellent agreement was obtained between the semi-analytical and numerical values, under all the interface pore conditions tested. The validity of  the numerical strategy used to compute the integrals in the semi-analytical method was therefore confirmed. From the physical point of view, Figure \ref{FigN1compare} presents three peaks due to the following waves: $Pf$ for the fast compressional wave ($t_{Pf}\approx0.21\;$ms), $F$ for the direct fluid wave ($t_{F}=0.33\;$ms), and $St$ for the pseudo-Stoneley wave ($t_{St}\approx0.51\;$ms). The influence of the pore conditions adopted at the fluid / porous interface can be clearly seen in Fig. \ref{FigN1pores}. Note that the direct fluid wave occurring in the acoustic pressure component (R1) was not affected by the pore conditions, contrary to what occurred with all the other peaks.


\subsection{Test 2: $N=4$ interfaces}\label{SecResN4}

The computational domain was the same here as in test 1 and the coordinates of the 4 interfaces were set as follows: $\alpha_0=0$ m, $\alpha_1=-0.1$ m, $\alpha_2=-0.15$ m, and $\alpha_3=-0.6$ m, with layer thicknesses ranging from 0.5 to 4 wavelengths of the fast compressional wave. Once again, $\Delta x=\Delta y=2.10^{-3}$ m were used for the coarse grid, and $\Delta x/5,\,\Delta y/5$ in the more highly refined grids surrounding each interface. 

\begin{figure}[htbp]
\begin{center}
\begin{tabular}{c}
\includegraphics[scale=0.60]{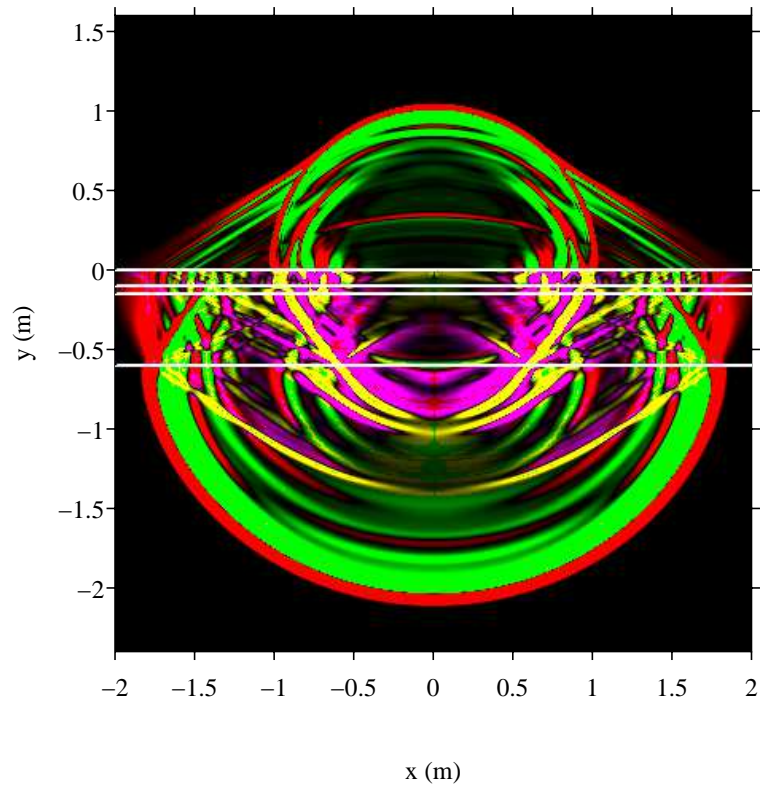} 
\end{tabular}
\end{center}
\vspace{-1cm}
\caption{$N=4$ interfaces involving open pores at the fluid/porous boundary. Snapshot of $p$ (in the fluid) and $\sigma_{xx}$ (in the poroelastic layers) at $t=0.69$ ms.}
\label{FigN4Carte}
\end{figure}

\begin{figure}[htbp]
\begin{center}
\begin{tabular}{cc}
$p$ (open pores)   & $\dot{u}_y$ (open pores)   \\
\includegraphics[scale=0.35]{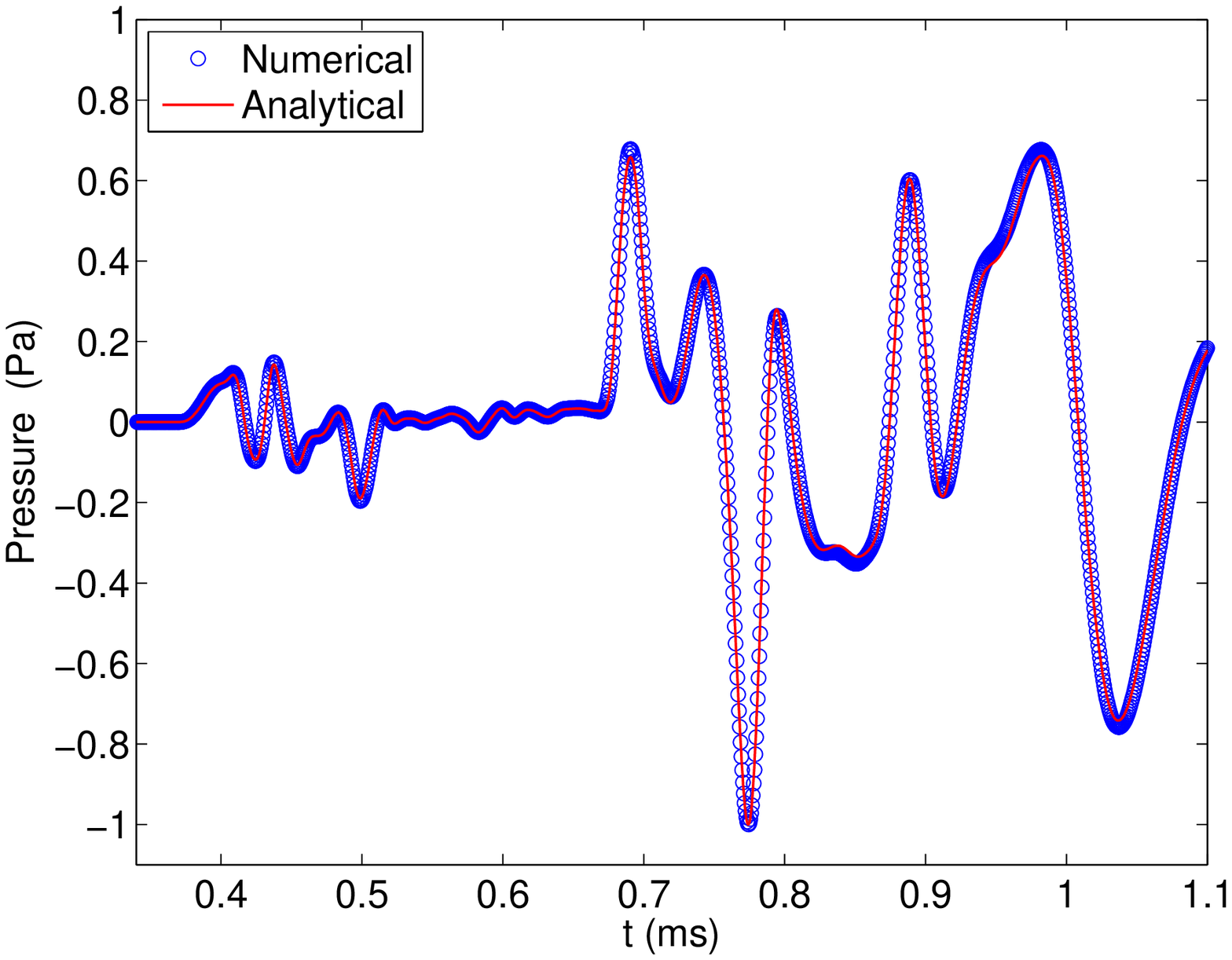} &
\includegraphics[scale=0.35]{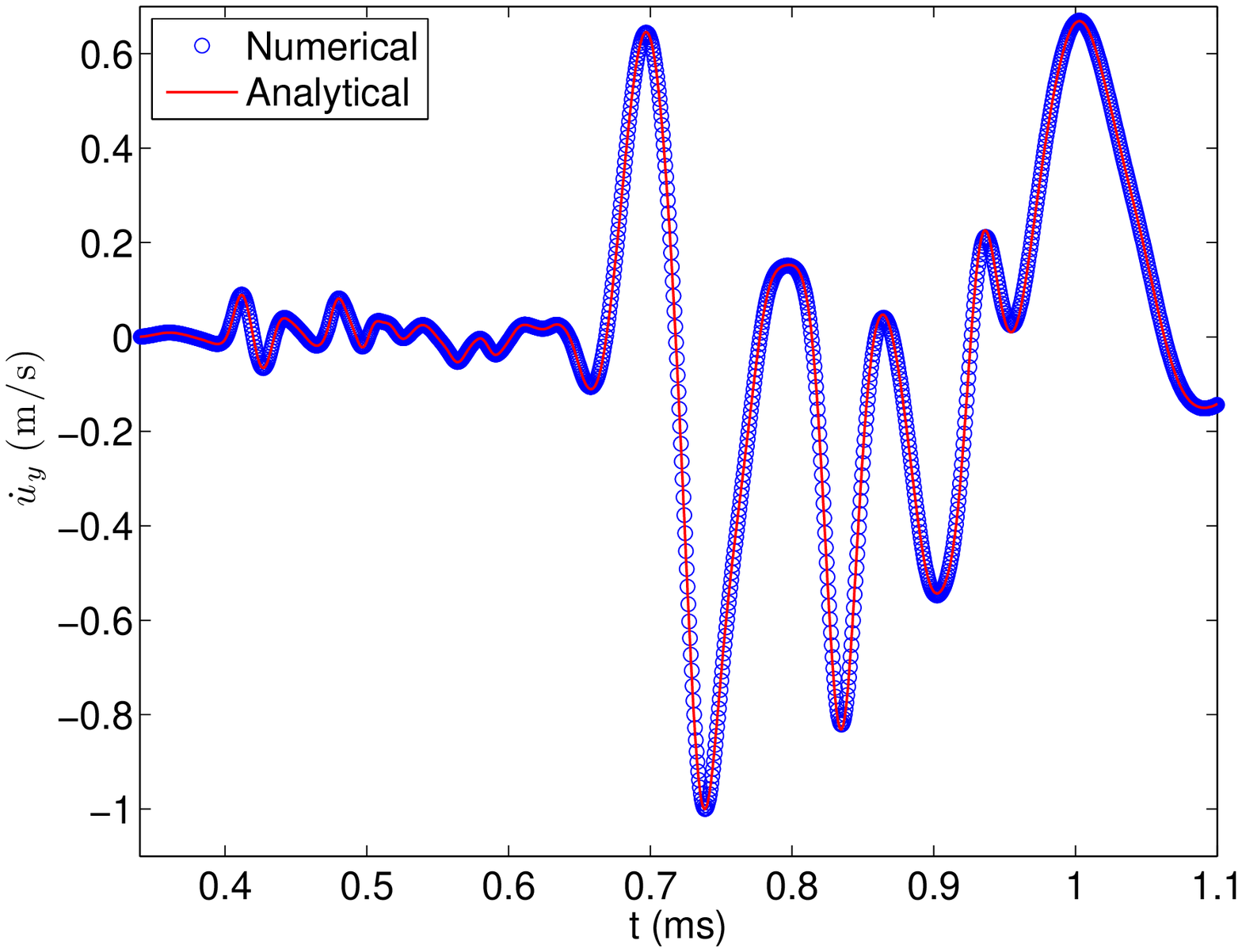}  \\ 
$p$ (imperfect pores)   & $\dot{u}_y$ (imperfect pores)   \\
\includegraphics[scale=0.35]{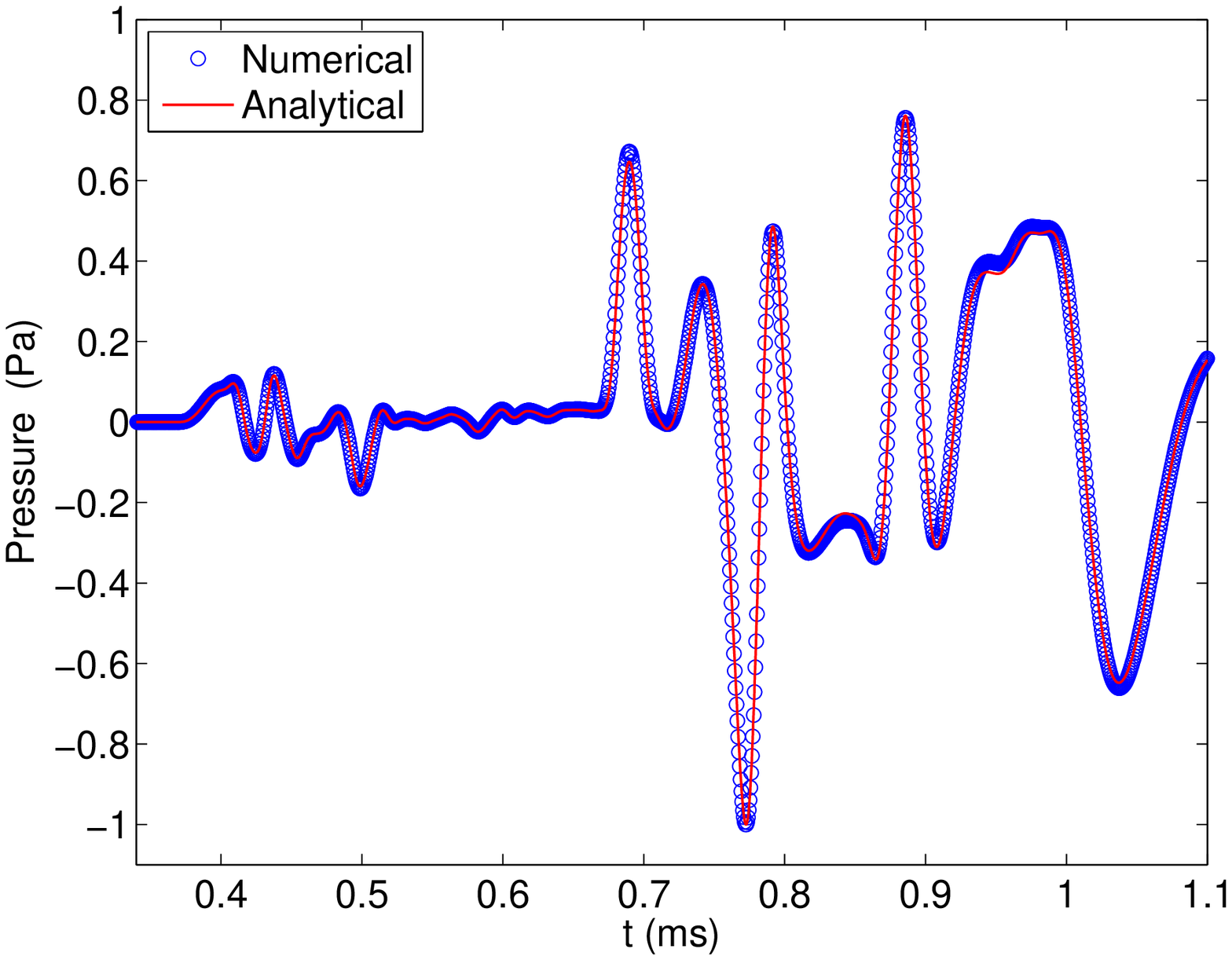} &
\includegraphics[scale=0.35]{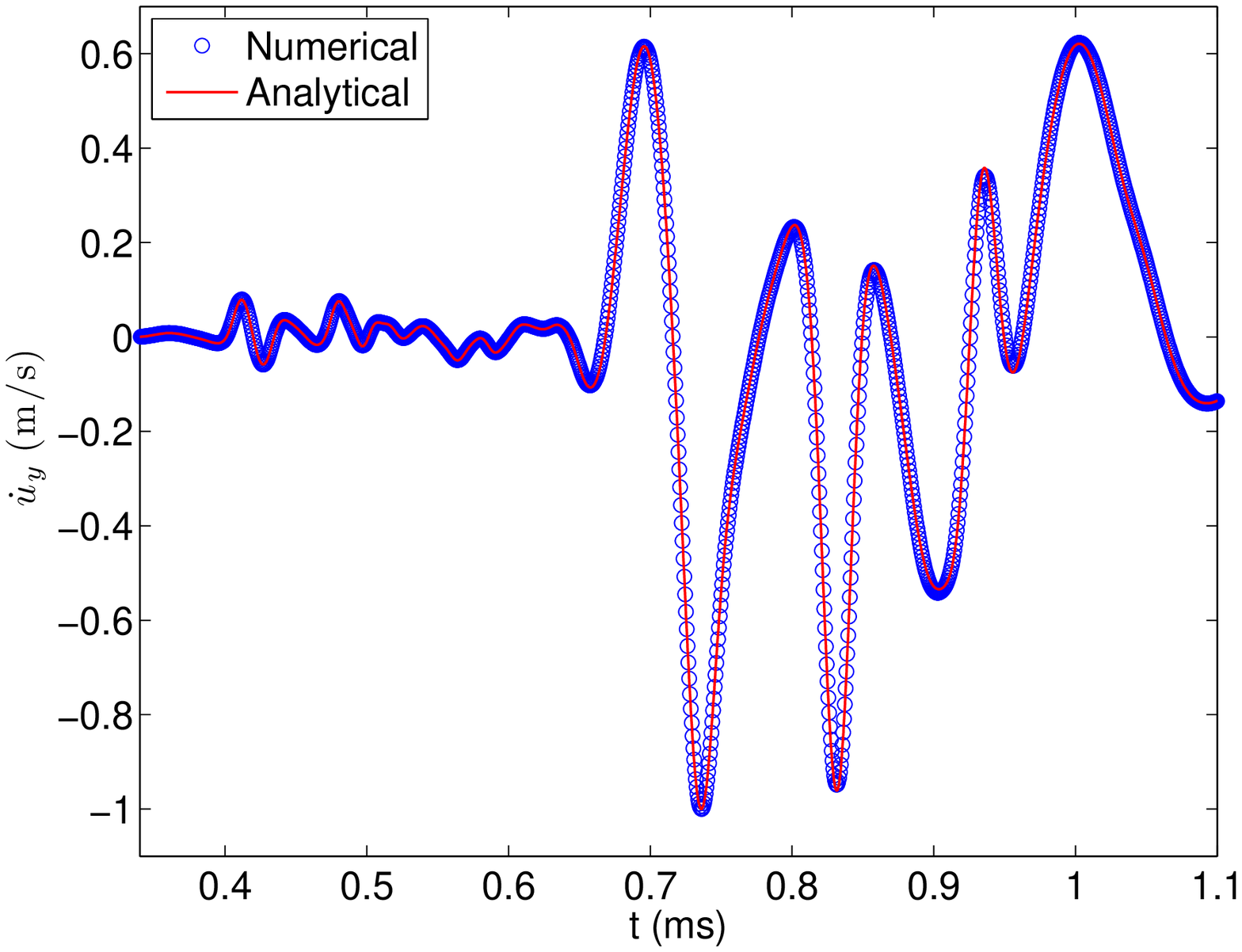}  \\ 
$p$ (sealed pores) & $\dot{u}_y$ (sealed pores) \\
\includegraphics[scale=0.35]{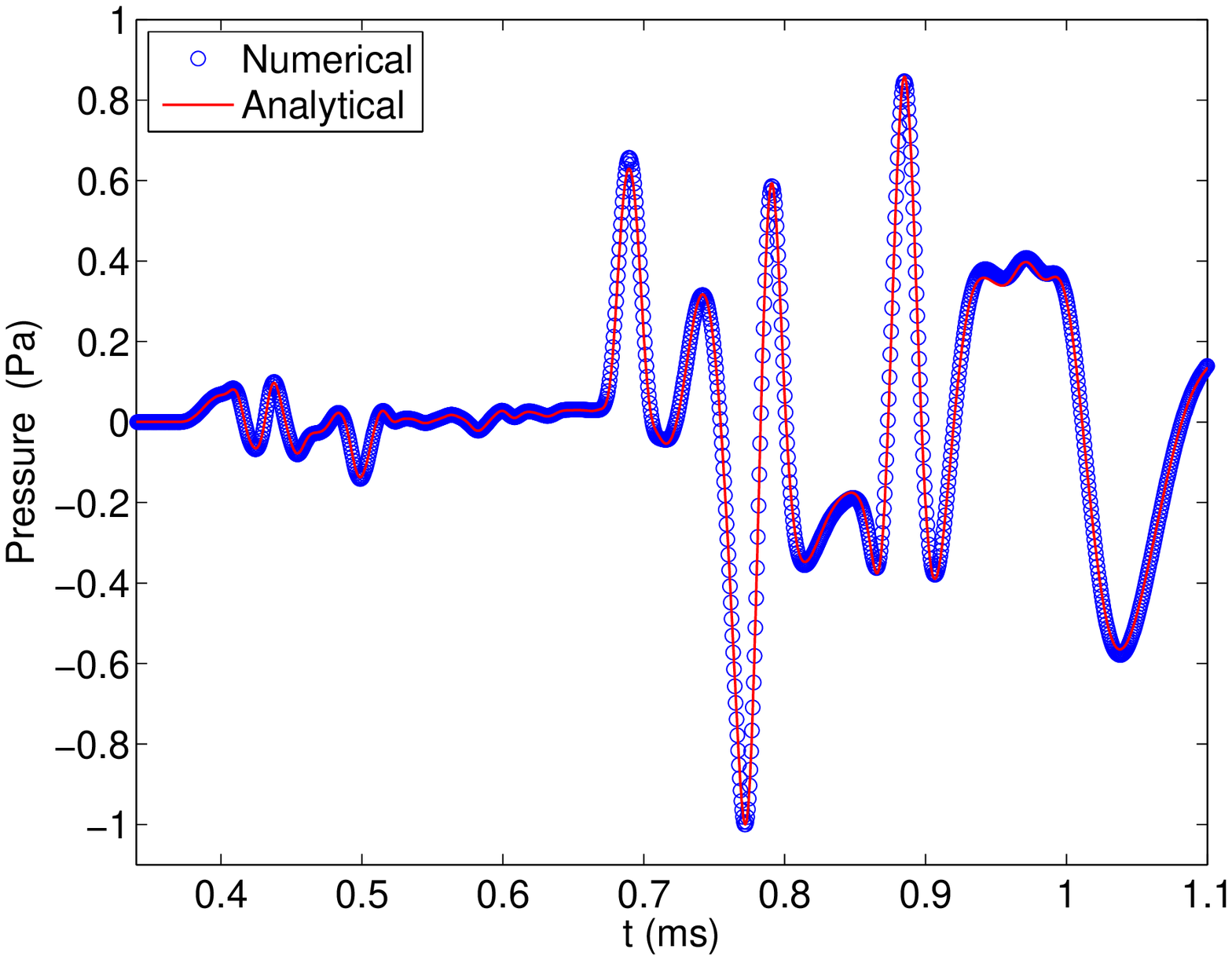}&
\includegraphics[scale=0.35]{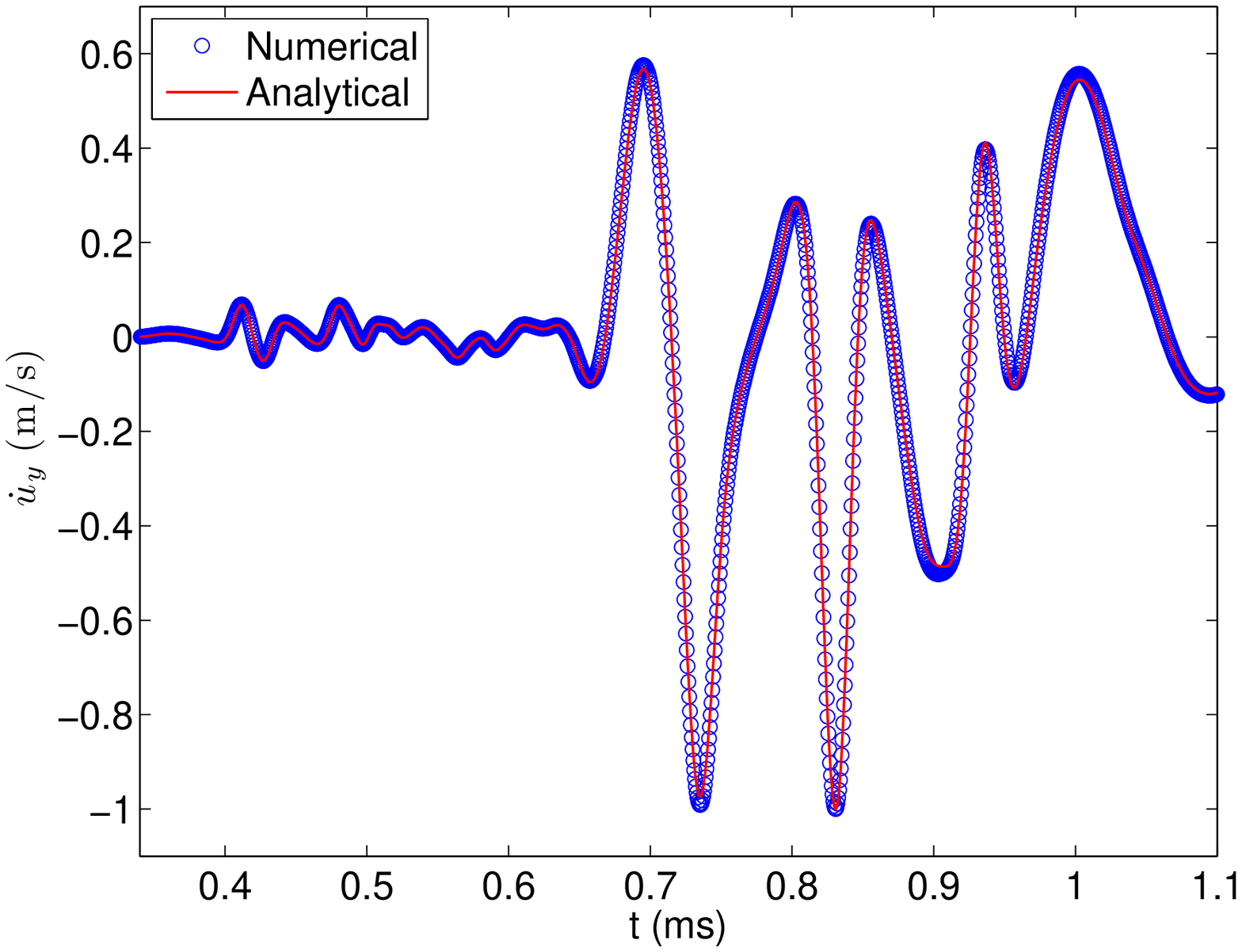} \\
\end{tabular}
\end{center}
\vspace{-.5cm}
\caption{$N=4$ interfaces, with open pores (top), imperfect pores (middle) and sealed pores (bottom). Time evolution of the acoustic pressure $p$ at recorder R3 $(x=1,\,y=0.02)$ m (left) and that of the vertical velocity $\dot{u}_y$ at recorder R4 $(x=1,\,y=-0.112)$ m (right).}
\label{FigN4Compare}
\end{figure}

The numerical solution is presented in Fig. \ref{FigN4Carte}, where complex interactions and wave conversions can be observed. The time evolution of the acoustic pressure at point R3 $(x=1,\,y=0.02)$ m and that of the vertical velocity $\dot{u}_y$ at point R4 $(x=1,\,y=-0.112)$ m are shown in Fig. \ref{FigN4Compare}. Once again, excellent agreement was observed between the semi-analytical and numerical values obtained. With the conditioning matrix technique, no oscillation occurs in the semi-analytical values, and the layers therefore do not need to be divided into sublayers, which makes the calculations more efficient \cite{MESGOUEZ09}.

\begin{figure}[htbp]
\begin{center}
\begin{tabular}{cc}
\includegraphics[scale=0.35]{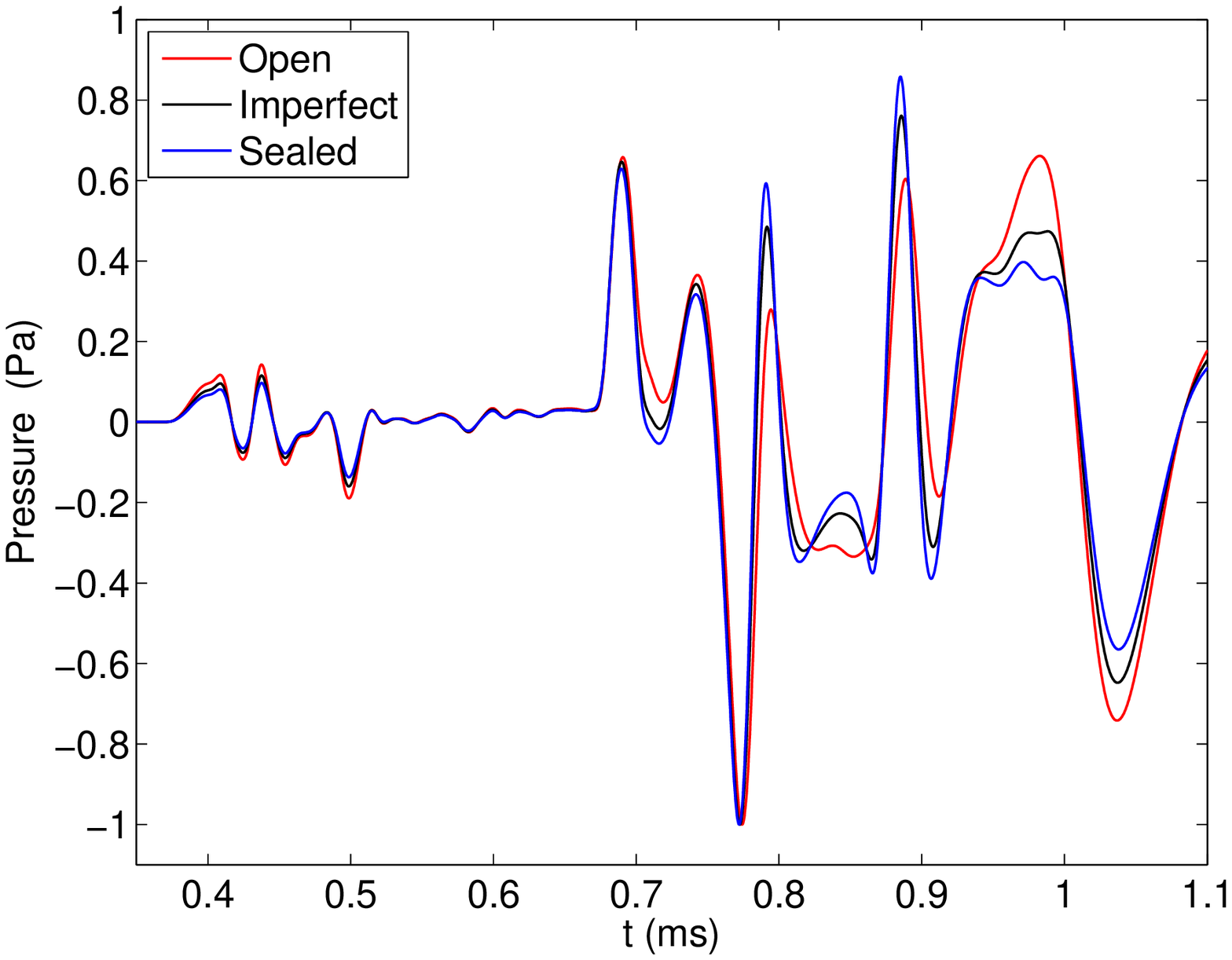} 
\includegraphics[scale=0.35]{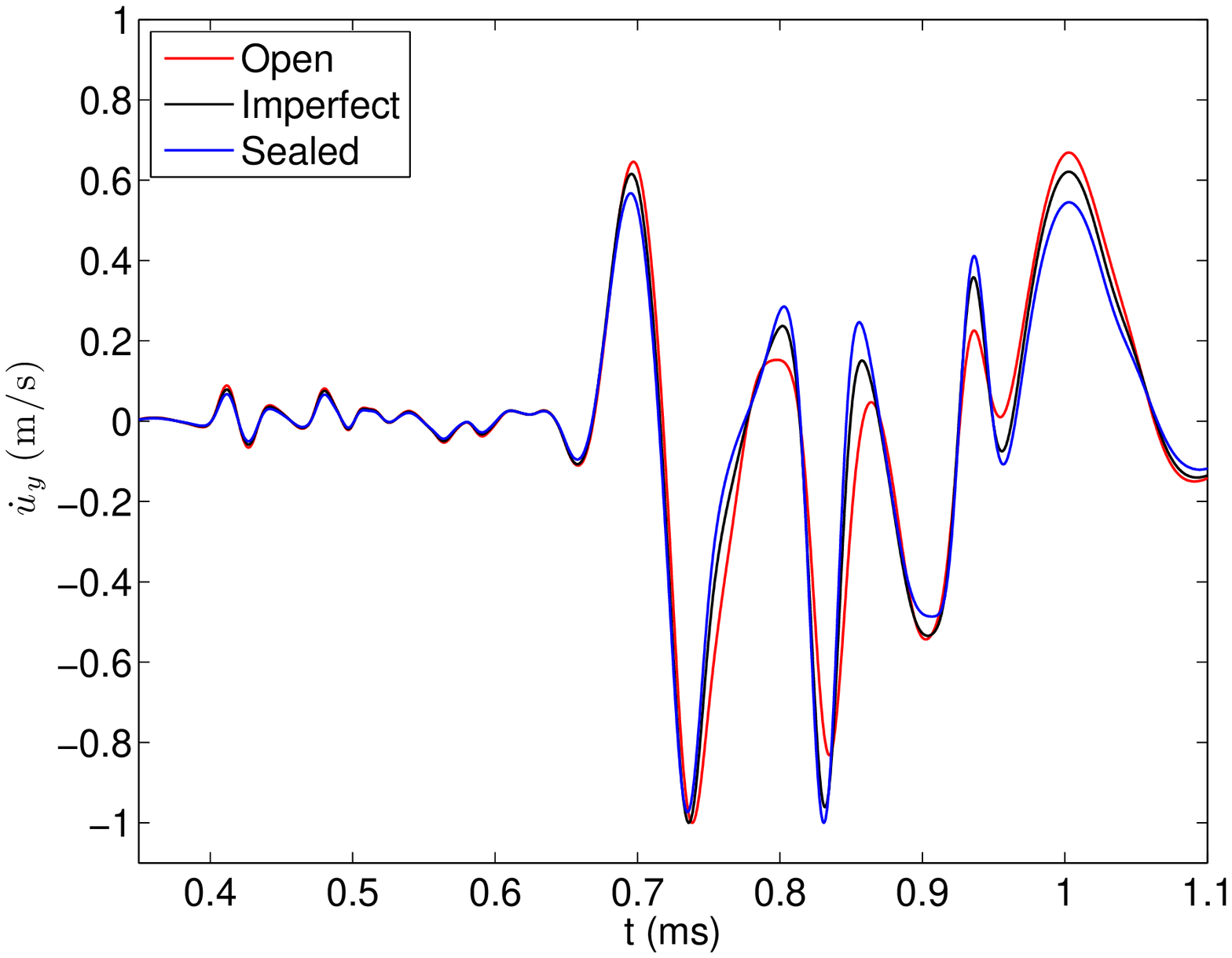} 
\end{tabular}
\end{center}
\vspace{-.5cm}
\caption{$N=4$ interfaces. Comparison between the different pore conditions: pressure (left) and vertical velocity (right). }
\label{FigN4toutPression}
\end{figure}

Differences between the pressures computed under open, imperfect and sealed conditions can be clearly seen in Fig. \ref{FigN4toutPression}, especially in the successive diffracted waves. This shows the need for an accurate means of treating the pore conditions whatever the numerical method used, even in the case of a single fluid / porous interface.  

\section{Conclusion}\label{SecConclu}

The propagation of transient waves in a stratified 2D fluid / poroelastic medium was studied here using two different approaches: a semi-analytical and a numerical approach. On the one hand, the analytical solution is fast, and can therefore be efficiently plugged into imaging codes \cite{LEFEUVE12, DEBARROS10}. This approach can only be used with academic geometries, however. On the other hand, the numerical solution can be used to handle much more complex geometries and media, but the full space-time grid is not suitable when the solution is sought only at some receivers. 
 
Both methods are also able to deal with various boundary conditions, which can have non-negligible effects on the various fields, as we have seen here. As far as we know, the case of imperfect pores has never been treated previously using a numerical approach. Since the assumptions underlying the present semi-analytical and numerical approaches are radically different, the comparisons between the results obtained constitute an authentic cross-check, and the results obtained therefore provide useful benchmarks for future researches. 

We are currently extending this study to include frequencies greater than Eq. (\ref{Fc}), using the JKD model \cite{JKD87}. The frequency correction can be straightforwardly incorporated into the analytical method, but the time-domain numerical modeling needs to be considerably adapted because of fractional derivatives involved \cite{HANYGA05}. Work is currently under way on these lines \cite{THESE_BLANC}.

\section*{Acknowledgements}
The authors wish to thank Prof. A. Wirgin and Dr. E. Ogam for their helpful comments, and Jessica Blanc for her careful reading of the manuscript.

\appendix

\section{Notations used along section \ref{SecAnal}}\label{SecAppendixMatrices}

The vertical components of the wavenumbers (\ref{Dispersion}), which are denoted by an $y$ index, satisfy

\begin{equation}
k_{yPj}^2=k_{Pj}^2-k_x^2,\qquad k_{yS}^2=k_S^2-k_x^2. \label{eq:wavenbs}
\end{equation}
To ensure the radiation condition, we choose $\Im (k_{yPj},\;k_{yS})\geq 0$. Then, the coefficients associated with compressional waves and the shear wave are 
\begin{equation}
\begin{array}{lll}
F_{Pj}&=&
\displaystyle
\frac{\textstyle (k_{yPj}^2+k_x^2)\beta\,m-\omega^2\,\rho_f}{\textstyle -(k_{yPj}^2+k_x^2)\,m+\omega^2 {\textstyle a_{\infty}\,\rho_f}/{\textstyle \phi}+i\omega\,{\textstyle \eta}/{\textstyle \kappa}},\quad j=f,s\,\mbox{ (fast, slow)},\\
G&=&
\displaystyle
-\frac{\textstyle \phi\,\rho_f\,\kappa\,\omega}{\textstyle a_{\infty}\,\rho_f\,\kappa\,\omega+i\,\eta\, \phi}.
\end{array}
\label{FG}
\end{equation}

\noindent
The matrices ${\bf Z}$, $\bf{S}^{T,R}$ and $\bf{Q}^{T,R}$ in system (\ref{parite}) are given by
\begin{equation}
{\bf Z} (h_k)={\bf diag} <\exp(i\,k_{yPf}\,h_k),\,\exp(i\,k_{yPs}\,h_k),\,\exp(i\,k_{yS}\,h_k)>,
\label{MatZ}
\end{equation}

where $\bf{diag}$ stands for the diagonal matrix,
\begin{equation}
\begin{array}{l}
\bf{S}^T=\left(
\begin{array}{ccc}
S_{11}^T & S_{12}^T & S_{13}^T \\
[4pt]
S_{21}^T & S_{22}^T & S_{23}^T \\
[4pt]
S_{31}^T & S_{32}^T & 0 
\end{array}
\right),
\quad
\bf{S}^R=\left(
\begin{array}{ccc}
-S_{11}^T & -S_{12}^T &  S_{13}^T \\
[4pt]
 S_{21}^T &  S_{22}^T & -S_{23}^T \\
 [4pt]
 S_{31}^T &  S_{32}^T &  0 
\end{array}
\right),\\
\\
S_{11}^T =  2\,\mu\,k_{yPf}\,k_x,\\
[4pt]
S_{12}^T =  2\,\mu\,k_{yPs}\,k_x,\\
[4pt]
S_{13}^T =  \mu\,(k_{yS}^2-k_x^2),\\
[4pt]
S_{21}^T =  i\,(-(k_{yPf}^2+k_x^2)\,(\lambda_0+m\,\beta^2+m\,\beta F_{Pf})-2\,\mu\,k_{yPf}^2),\\
[4pt]
S_{22}^T =  i\,(-(k_{yPs}^2+k_x^2)\,(\lambda_0+m\,\beta^2+m\,\beta F_{Ps})-2\,\mu\,k_{yPs}^2),\\
[4pt]
S_{23}^T =  2\,i\,\mu\,k_{yS}\,k_x,\\
[4pt]
S_{31}^T = -i\,m\,(k_{yPf}^2+k_x^2)\,(F_{Pf}+\beta),\\
[4pt]
S_{32}^T = -i\,m\,(k_{yPs}^2+k_x^2)\,(F_{Ps}+\beta),
\end{array}
\label{MatS}
\end{equation}
and
\begin{equation}
\bf{Q}^T=
\left(
\begin{array}{ccc}
i\,k_x & i\,k_x & i\,k_{yS} \\
[4pt]
k_{yPf} & k_{yPs} & -k_x \\
[4pt]
k_{yPf} F_{Pf} & k_{yPs} F_{Ps} & -k_x\,G
\end{array}
\right),
\quad
\bf{Q}^R=\left(
\begin{array}{ccc}
 Q_{11}^T &  Q_{12}^T & -Q_{13}^T \\
 [4pt]
-Q_{21}^T & -Q_{22}^T &  Q_{23}^T \\
[4pt]
-Q_{31}^T & -Q_{32}^T &  Q_{33}
\end{array}
\right).
\label{MatQ}
\end{equation}

\bibliographystyle{model1-num-names}
\bibliography{<biblio>}

\end{document}